\documentclass[aps,prd,twoside,twocolumn,nofootinbib,showpacs,floatfix]{revtex4-1}
\usepackage{amssymb}
\usepackage{amsmath,amssymb}
\usepackage{graphicx,bm}
\usepackage{slashed}
\usepackage{times}
\usepackage{epstopdf}
\usepackage{ulem} 
\usepackage[usenames]{color}
\usepackage{float}
\usepackage{subfigure}
\usepackage{subfigure}
\usepackage{rotating}
\usepackage{color}
\usepackage{multirow}
\usepackage{dcolumn}
\usepackage{overpic}
\usepackage{booktabs}
\usepackage{appendix}
\usepackage{makecell}
\usepackage{diagbox}

\renewcommand\sout{\bgroup \color{red} \ULdepth=-.5ex \ULset}

\usepackage[colorlinks, citecolor=blue,anchorcolor=red,menucolor=red,linkcolor=blue,filecolor=red,runcolor=red,urlcolor=blue,frenchlinks=red]{hyperref}
\allowdisplaybreaks[4]
\begin{document}
\title{From the $P^{N}_{\psi}$/$P^{\Lambda}_{\psi s}$ to $\bar{T}^f_{cc}$: symmetry analysis to the interactions of the $(\bar{c}q)(\bar{c}q)$/$(ccq)(\bar{c}q)$/$(ccq)(ccq)$ di-hadron systems}
\author{Kan Chen}\email{chenk10@nwu.edu.cn}
\affiliation{School of Physics, Northwest University, Xi'an 710127, China}
\affiliation{Shaanxi Key Laboratory for theoretical Physics Frontiers, Xi'an 710127, China}
\affiliation{Institute of Modern Physics, Northwest University, Xi'an 710127, China}
\affiliation{Peng Huanwu Center for Fundamental Theory, Xi'an 710127, China}
\author{Bo Wang}\email{wangbo@hbu.edu.cn}
\affiliation{College of Physics Science \& Technology, Hebei University, Baoding 071002, China}
\affiliation{Hebei Key Laboratory of High-precision Computation and Application of Quantum Field Theory, Baoding, 071002, China}
\affiliation{Hebei Research Center of the Basic Discipline for Computational Physics, Baoding, 071002, China}

\begin{abstract}
We investigate the interactions of the $(\bar{c}q)(\bar{c}q)$/$(ccq)(\bar{c}q)$/$(ccq)(ccq)$ di-hadron systems based on a contact lagrangian possessing the SU(3) flavor and SU(2) spin symmetries. Under the assumptions of two scenarios for the $J^P$ quantum numbers of the $P_{\psi}^N(4440)$ and $P_{\psi}^N(4457)$ states, we obtain the parameters ($\tilde{g}_s$, $\tilde{g}_a$) introduced from this contact lagrangian. Then we include the SU(3) breaking effect by introducing a factor $g_x$, this quantity can be further constrained by the experimental mass of the $P_{\psi s}^\Lambda(4338)$ state. We can reproduce the mass of the $T^f_{cc}(3875)$ state with the parameters extracted from the observed $P_{\psi}^N$ states, this consistency indicates a unified description of the di-hadron molecular states composed of two heavy-light hadrons. With the same parameters, we discuss the possible mass spectra of the $\bar{T}_{cc}^f$/$P_{\psi c}^\Lambda$/$H_{\Omega_{ccc}c}^\Lambda$ systems. Then we proceed to discuss the existences of the $\bar{T}_{cc\bar{s}}^\theta$/$P_{\psi cs}^N$/$H_{\Omega_{ccc}cs}^N$ states by investigating the SU(3) breaking effects. Our results show that the states in the $\bar{T}_{cc\bar{s}}^\theta$/$P_{\psi cs}^N$ systems can hardly form bound states, while the states in the $H_{\Omega_{ccc}cs}^N$ system can form bound states due to their larger reduced masses. 
\end{abstract}
\maketitle

\vspace{2cm}
\section{Introduction}\label{Introduction}


The discoveries of the $P_{\psi}^N$ \cite{LHCb:2015yax,LHCb:2019kea}, $P_{\psi s}^\Lambda$ \cite{LHCb:2020jpq,LHCb:2022ogu}, and $T^f_{cc}(3875)$ \cite{LHCb:2021auc,LHCb:2021vvq} states\footnote{We adopt the nomenclature proposed by the LHCb collaboration \cite{Gershon:2022xnn} throughout this paper.} significantly expand our understanding of exotic hadronic states. To elucidate their underlying structures, several theoretical frameworks have been proposed, including the molecular interpretations, compact pentaquark/tetraquark states, hadro-charmonium states, and kinematical effects \cite{Chen:2016qju,Lebed:2016hpi,Oset:2016lyh,Esposito:2016noz,Chen:2016spr,Richard:2016eis,Dong:2017gaw,Guo:2017jvc,Olsen:2017bmm,Ali:2017jda,Karliner:2017qhf,Guo:2019twa,Brambilla:2019esw,Liu:2019zoy,Chen:2022asf,Meng:2022ozq,Liu:2024uxn}. The proximity of the masses of the $P_{\psi}^N$, $P_{\psi s}^\Lambda$, and $T^f_{cc}(3875)$ states to their respective di-hadron thresholds strongly suggests their molecular configurations. Consequently, the molecular picture has emerged as the most widely accepted interpretation for these manifestly exotic states.

With the accumulation of experimental data, we can anticipate the identification of more molecular candidates in the near future. In the molecular picture, the residual strong interaction, primarily driven by the exchange of light mesons, facilitates the binding of a di-hadron system. Specifically, in an $S$-wave di-hadron system composed of two ground-state heavy-light hadrons, the characteristic of the residual strong interaction is contingent upon the correlations between the light quark components in each hadron. Consequently, the interactions among different di-hadron systems can be interconnected through SU(3) flavor and SU(2) spin symmetries, based on the flavor and spin structures arising from the light degrees of freedom (d.o.f.) within each di-hadron system.

The light quark components in the di-hadron system determine the property of the interaction among these two hadrons, while the heavy quark components are important in stabilizing an attractive di-hadron system. For example, in our previous work \cite{Chen:2024tuu}, by considering the heavy diquark-antiquark symmetry (HDAS) \cite{Savage:1990di,Guo:2013sya,Liu:2018zzu,Yang:2020nrt}, we related the $\bar{D}^{(*)}$ and $\bar{D}^{(*)}_s$ mesons to the $\Xi_{cc}^{(*)}$ and $\Omega_{cc}^{(*)}$ baryons, respectively. With the same attractive force, the di-baryon $H_{\Omega_{ccc}}^N$ and $H_{\Omega_{ccc}s}^{\Lambda}$ systems bind more deeper than that of the baryon-meson (B-M) $P_{\psi}^N$ and $P_{\psi s}^\Lambda$ systems, respectively.

Thus, for an $S$-wave di-hadron system composed of two ground heavy-light hadrons, the light quark components and heavy quark components play distinct roles in forming molecules. The SU(3) flavor and SU(2) spin symmetries relate the interactions of the di-hadron systems with different light quark components, i.e., relating the $P_{\psi}^N$ ($cnn+\bar{c}n$) ($n=u$, $d$) and $P_{\psi s}^\Lambda$ ($cnn+\bar{c}s$ or $cns+\bar{c}n$) states to the $\bar{T}^f_{cc}$ ($\bar{c}n+\bar{c}n$) and $\bar{T}^\theta_{cc\bar{s}}$ ($\bar{c}n+\bar{c}s$) states, respectively. While the HDAS relates the interactions of the di-hadron systems with different heavy quark components, i.e., relating the $\bar{T}_{cc}^f$ and $\bar{T}_{cc\bar{s}}^\theta$ states to the $P_{\psi c}^\Lambda$ ($ccn+\bar{c}n$) and $P_{\psi cs}^N$ ($ccn+\bar{c}s$ or $ccs+\bar{c}n$) states, or relating the $\bar{T}_{cc}^f$ and $\bar{T}_{cc\bar{s}}^\theta$ states to the $H_{\Omega_{ccc}c}^{f}$ ($ccn-ccn$) and $H_{\Omega_{ccc}cs}^{\theta}$ ($ccn-ccs$) states.

In this work, we investigate the interactions of the $\bar{T}_{cc}^f$/$\bar{T}_{cc\bar{s}}^\theta$, $P_{\psi c}^\Lambda$/$P_{\psi cs}^N$, and $H_{\Omega_{ccc}c}^\Lambda$/$H_{\Omega_{ccc}cs}^N$ systems from a symmetric perspective. These three sets of systems have identical light quarks, and they are expected to have very similar interactions. Among these systems, the $\bar{T}_{cc}^f$ system attracted most attentions. Based on the molecule or compact tetraquark assumptions, the properties of the $\bar{T}_{cc}^f$ states are widely discussed in a large amount of theoretical works \cite{Meng:2021jnw,Wang:2022jop,Wang:2023hpp,Ling:2021bir,Agaev:2021vur,Maiani:2022qze,Lin:2022wmj,Cheng:2022qcm,Achasov:2022onn,Padmanath:2022cvl,Agaev:2022ast,Yang:2019itm,Deng:2021gnb,Santowsky:2021bhy,Du:2021zzh,Baru:2021ldu,Albaladejo:2021vln,Ren:2021dsi,Xin:2021wcr,Guo:2021yws,Qin:2020zlg,Meng:2023jqk,Lebed:2024zrp,Ortega:2024epk,Whyte:2024ihh,Sun:2024wxz}. As the strange partners of $\bar{T}_{cc}^f$ states, the existences, productions, possible mass spectra, and other relevant topics of the $\bar{T}_{cc\bar{s}}^\theta$ states have also been discussed in many literatures \cite{Qin:2020zlg,Junnarkar:2018twb,Wang:2024vjc,Mutuk:2024vzv,Li:2023wug,Mutuk:2023oyz,Meng:2023jqk,Ma:2023int,Li:2023hpk,Albuquerque:2023rrf,Azizi:2023gzv,Praszalowicz:2022sqx,Chen:2022ros,Wang:2022clw,Kim:2022mpa,Karliner:2021wju,Guo:2021yws}. Besides, the triple-charm $P_{\psi c}^\Lambda$ molecular pentaquarks have been investigated within the framework of one-boson-exchange model \cite{Chen:2017jjn,Asanuma:2023atv}.  However, to our knowledge, theoretical researches on the interactions of the $P_{\psi cs}^N$, $H_{\Omega_{ccc}c}^\Lambda$, $H_{\Omega_{ccc}cs}^N$ systems is missing at present. This work is devoted to study and relate the interactions of the $\bar{T}_{cc}^f$/$\bar{T}_{cc\bar{s}}^\theta$, $P_{\psi c}^\Lambda$/$P_{\psi cs}^N$, and $H_{\Omega_{ccc}c}^\Lambda$/$H_{\Omega_{ccc}cs}^N$ systems in a unified framework.

This work is organised as follows, we present our framework in Sec. \ref{Framework}, then we present our numerical results and discussions for the $\bar{T}_{cc}^f$/$\bar{T}_{cc\bar{s}}^\theta$, $P_{\psi c}^\Lambda$/$P_{\psi cs}^N$, and $H_{\Omega_{ccc}c}^\Lambda$/$H_{\Omega_{ccc}cs}^N$ systems in Sec. \ref{Numerical}. Sec. \ref{summary} is devoted to a summary.

\section{Framework}\label{Framework}
Firstly, we list the considered multi-heavy di-hadron systems with the strangeness numbers 0 and $-1$ in Table \ref{thresholds}. We also list their corresponding thresholds in Table \ref{thresholds}. Here, we adopt the experimental mass of the $\Xi_{cc}$ baryon from the LHCb collaboration \cite{LHCb:2017iph,LHCb:2022rpd}, and adopt the theoretical masses of the $\Xi_{cc}^*$ and $\Omega_{cc}^{(*)}$ baryons calculated within a relativistic quark model  \cite{Ebert:2002ig}.

\begin{table}[htbp]
\renewcommand\arraystretch{1.5}
\caption{The considered multi-heavy $(\bar{c}q)(\bar{c}q)$/$(ccq)(\bar{c}q)$/$(ccq)(ccq)$ di-hadron systems with the strangeness numbers 0 and $-1$. To list the thresholds of the considered systems, we adopt the experimental mass of the $\Xi_{cc}^{++}$ baryon measured from the LHCb collaboration \cite{LHCb:2017iph,LHCb:2022rpd}, and adopt the theoretical masses of the $\Xi_{cc}^*$ and $\Omega_{cc}^{(*)}$ baryons calculated from Ref. \cite{Ebert:2002ig}. All values are in units of MeV.}
\begin{tabular}{c|cccccccccccccc}
\toprule[0.8pt]
\multirow{6}{*}{$S=0$}&$\bar{D}\bar{D}$&\multicolumn{2}{c}{$\bar{D}^*\bar{D}$}&$\bar{D}^*\bar{D}^*$\\
&3734.5&\multicolumn{2}{c}{3875.6}&4017.1\\
&$\Xi_{cc}\bar{D}$&$\Xi^*_{cc}\bar{D}$&$\Xi_{cc}\bar{D}^*$&$\Xi_{cc}^{*}\bar{D}^*$\\
&5488.6&5594.2&5630.0&5735.6\\
&$\Xi_{cc}\Xi_{cc}$&\multicolumn{2}{c}{$\Xi^*_{cc}\Xi_{cc}$}&$\Xi_{cc}^*\Xi_{cc}^*$\\
&7242.8&\multicolumn{2}{c}{7348.4}&7454.0\\
\hline
\multirow{8}{*}{$S=-1$}&$\bar{D}\bar{D}_s$&$\bar{D}^*\bar{D}_s$&$\bar{D}\bar{D}^*_s$&$\bar{D}^*\bar{D}_{s}^*$\\
&3836.2&3977.6&3979.4&4120.8\\
&$\Xi_{cc}\bar{D}_s$&$\Xi^*_{cc}\bar{D}_s$&$\Xi_{cc}\bar{D}^*_s$&$\Xi_{cc}^*\bar{D}_s^*$\\
&5590.4&5696.0&5733.6&5839.2\\
&$\Omega_{cc}\bar{D}$&$\Omega^*_{cc}\bar{D}$&$\Omega_{cc}\bar{D}^*$&$\Omega_{cc}^{*}\bar{D}^*$\\
&5645.2&5739.2&5786.6&5880.6\\
&$\Xi_{cc}\Omega_{cc}$&$\Xi^*_{cc}\Omega_{cc}$&$\Xi_{cc}\Omega^*_{cc}$&$\Xi_{cc}^*\Omega_{cc}^*$\\
&7399.4&7505.0&7493.4&7599.0\\
\bottomrule[0.8pt]
\end{tabular}\label{thresholds}
\end{table}
\subsection{The flavor-spin wave functions for all the considered systems}
To describe the flavor-spin wave functions of the heavy-light di-hadron systems listed in Table \ref{thresholds}, we introduce three types of notations to denote the total wave functions of the considered di-hadron states. Explicitly, we use the
\begin{eqnarray}
\left|\left[H_1H_2\right]_J^I\right\rangle&=&\sum_{m_{I_1}m_{I_2}}C_{I_1,m_{I_1};I_2,m_{I_2}}^{I,I_z}\phi_{I_1,m_{I_1}}^{H_1}\phi_{I_2,m_{I_2}}^{H_2}\nonumber\\
&&\otimes\sum_{m_{S_1},m_{S_2}}C_{S_1,m_{S_1};S_2,m_{S_2}}^{J,J_z}\phi_{S_1,m_{S_1}}^{H_1}\phi_{S_2,m_{S_2}}^{H_2}\nonumber\label{wave1}\\
\end{eqnarray}
to denote the $S=0$ di-hadron states with a definite total isospin $I$ and a total angular momentum $J$. We use the
\begin{eqnarray}
\left|\left[H_1H_2\right]_J^{\frac{1}{2}(\pm)}\right\rangle&=&\sqrt{\frac{1}{2}}\left(\phi_{I_1,m_{I_1}}^{H_1}\phi_{I_2,m_{I_2}}^{H_2}\pm \phi_{I_2,m_{I_2}}^{H_2}\phi_{I_1,m_{I_1}}^{H_1}\right)\nonumber\\
&&\otimes\sum_{m_{S_1},m_{S_2}}C_{S_1,m_{S_1};S_2,m_{S_2}}^{J,J_z}\phi_{S_1,m_{S_1}}^{H_1}\phi_{S_2,m_{S_2}}^{H_2}\nonumber\label{wave2}\\
\end{eqnarray}
to denote the $S=-1$ di-hadron states that have a definite symmetry in the flavor space and with a total angular momentum $J$. Finally, we use the
\begin{eqnarray}
\left|\left[H_1H_2\right]_J\right\rangle&=&\phi_{I_1,m_{I_1}}^{H_1}\phi_{I_2,m_{I_2}}^{H_2}\nonumber\\
&&\otimes\sum_{m_{S_1},m_{S_2}}C_{S_1,m_{S_1};S_2,m_{S_2}}^{J,J_z}\phi_{S_1,m_{S_1}}^{H_1}\phi_{S_2,m_{S_2}}^{H_2}\nonumber\label{wave3}\\
\end{eqnarray}
to denote the $H_1H_2$ states that do not have a definite symmetry in the flavor space but with a definite total angular momentum $J$.
The $C_{I_1,m_{I_1};I_2,m_{I_2}}^{I,I_z}$ and $C_{S_1,m_{S_1};S_2,m_{S_2}}^{J,J_z}$ are the Clebsch-Gordan coefficients. The $\phi_{I_1,m_{I_1}}^{H_1}$ ($\phi_{I_2,m_{I_2}}^{H_2}$) and $\phi_{S_1,m_{S_1}}^{H_1}$ ($\phi_{S_2,m_{S_2}}^{H_2}$) are the flavor and spin wave functions of the $H_1$ ($H_2$) hadron, respectively.
\subsection{The effective potentials}
To describe the interactions of the considered di-hadron systems listed in Table \ref{thresholds}, we introduce the following leading order contact Lagrangian \cite{Wang:2019nvm,Wang:2020dhf,Chen:2021cfl,Chen:2021spf}
\begin{eqnarray}
\mathcal{L}=g_s\bar{q}\mathcal{S}q+g_a\bar{q}\gamma_\mu\gamma^5\mathcal{A}^\mu q,\label{Lagrangian}
\end{eqnarray}
where the $\mathcal{S}$ and $\mathcal{A}^\mu$ are the fictitious scalar and axial-vector fields, respectively. The redefined parameters are $\tilde{g}_s=g_s^2/m_{\mathcal{S}}^2$ and $\tilde{g}_a=g_a^2/m_{\mathcal{A}}^2$, where the $m_{\mathcal{S}}$ and $m_{\mathcal{A}}$ are the masses of the scalar and axial-vector light mesons, respectively.
Then the effective potential derived from Eq. (\ref{Lagrangian}) is
\begin{eqnarray}
V=\tilde{g}_s\bm{\lambda}_1\cdot\bm{\lambda}_2+\tilde{g}_a\bm{\lambda}_1\cdot\bm{\lambda}_2\bm{\sigma}_1\cdot\bm{\sigma}_2.
\end{eqnarray}
The $\bm{\lambda}_{1(2)}$ and $\bm{\sigma}_{1(2)}$ are the Gell-Mann and Pauli matrices in the flavor and spin spaces, respectively.
We can obtain the effective potential for a $[H_1H_2]_J^I$ (the same for the $[H_1H_2]_J^{\frac{1}{2}(\pm)}$ or $[H_1H_2]_J$ case) di-hadron state as
\begin{eqnarray}
V_{|[H_1H_2]_J^I\rangle}=\left\langle\left[H_1H_2\right]_J^I\left|V\right|\left[H_1H_2\right]_J^I\right\rangle.
\end{eqnarray}
Here, we need to emphasis that the operators $\bm{\lambda}_1\cdot\bm{\lambda}_2$ and $\bm{\sigma}_1\cdot\bm{\sigma}_2$ only act on the light quarks of the total wave function $|[H_1H_2]_J^I\rangle$ (the same for the $[H_1H_2]_J^{\frac{1}{2}(\pm)}$ or $[H_1H_2]_J$ case) in the flavor and spin spaces, respectively. Then we can extract the flavor ($\langle\mathcal{O}_{\text{tot}}^\text{f}\rangle$) and flavor-spin ($\langle\mathcal{O}_{\text{tot}}^{\text{fs}}\rangle$) matrix elements by calculating the
\begin{eqnarray}
\langle\mathcal{O}_{\text{tot}}^{\text{f}}\rangle=\left\langle\left[H_1H_2\right]_J^I\left|\bm{\lambda}_1\cdot\bm{\lambda}_2\right|\left[H_1H_2\right]_J^I\right\rangle\label{flavorop}
\end{eqnarray}
and
\begin{eqnarray}
\langle\mathcal{O}_{\text{tot}}^{\text{fs}}\rangle=\left\langle\left[H_1H_2\right]_J^I\left|\bm{\lambda}_1\cdot\bm{\lambda}_2\bm{\sigma}_1\cdot\bm{\sigma}_2\right|\left[H_1H_2\right]_J^I\right\rangle,
\end{eqnarray}
respectively. Note that the operators $\mathcal{O}^{\text{f}}_{\text{tot}}$ and $\mathcal{O}^{\text{fs}}_{\text{tot}}$ can be expanded as
\begin{eqnarray}
\mathcal{O}^\text{f}_{\text{tot}}=\mathcal{O}^{\text{f}}_1+\mathcal{O}^{\text{f}}_2+\mathcal{O}^{\text{f}}_3,\\
\mathcal{O}^\text{fs}_{\text{tot}}=\mathcal{O}^{\text{fs}}_1+\mathcal{O}^{\text{fs}}_2+\mathcal{O}^{\text{fs}}_3,
\end{eqnarray}
with
\begin{eqnarray}
\mathcal{O}_1^{\text{f}}&=&\sum_{i=1}^3\lambda_1^i\lambda_2^i,\quad\mathcal{O}_1^{\text{fs}}=\sum_{i=1}^3\lambda_1^i\lambda_2^i\bm{\sigma}_1\cdot\bm{\sigma}_2,\nonumber\\
\mathcal{O}_2^{\text{f}}&=&\sum_{j=4}^7\lambda_1^j\lambda_2^j,\quad\mathcal{O}_2^{\text{fs}}=\sum_{j=4}^7\lambda_1^j\lambda_2^j\bm{\sigma}_1\cdot\bm{\sigma}_2,\nonumber\\
\mathcal{O}_3^{\text{f}}&=&\lambda_1^8\lambda_2^8,\,\,\,\,\qquad\mathcal{O}_3^{\text{fs}}=\lambda_1^8\lambda_2^8\bm{\sigma}_1\cdot\bm{\sigma}_2.\nonumber
\end{eqnarray}
Thus, the matrix elements $\langle\mathcal{O}_1^{\text{f}}\rangle$ ($\langle\mathcal{O}_1^{\text{fs}}\rangle$), $\langle\mathcal{O}_2^{\text{f}}\rangle$ ($\langle\mathcal{O}_2^{\text{fs}}\rangle$), and $\langle\mathcal{O}_3^{\text{f}}\rangle$ ($\langle\mathcal{O}_3^{\text{fs}}\rangle$) represent the proportions of the exchanges from the isospin triplet, doublet, and singlet scalar (axial-vector) mesons, respectively.

In the SU(3) limit, the scalar (axial-vector) coupling parameters $\tilde{g}_s$ ($\tilde{g}_a$) for the exchanges of the isospin triplet, doublet, and singlet light mesons are the same. However, to take into account the SU(3) breaking effect, in Ref. \cite{Chen:2022wkh}, we suggested that the masses of the exchanged isospin doublet light strange mesons are heavier than that of the exchanged isospin triplet non-strange light mesons, in the non-relativistic limit, the propagator from the exchange of isospin doublet light meson is suppressed by the factor $1/m_{\text{ex}}^2$, where the $m_{\text{ex}}$ is the mass of the exchanged strange meson. Thus, we collectively multiply a SU(3) breaking factor $g_x$ in front of the matrix elements $\langle\mathcal{O}_2^{\text{f}}\rangle$ and $\langle\mathcal{O}_2^{\text{fs}}\rangle$ to suppress the contributions from the exchanges of the isospin doublet scalar and axial-vector light mesons, respectively.

To find the bound state solutions of the considered systems, we encounter both the single-channel and two-channel Lippmann-Schwinger (LS) equations. Specifically, in the single-channel case, we search for the bound state solutions by solving the following LS equation
\begin{eqnarray}
T(E)=V+VG(E)T(E).
\end{eqnarray}
Correspondingly, in the two-channel case, the corresponding LS equation has the following matrix form
\begin{eqnarray}
\mathbb{T}(E)=\mathbb{V}+\mathbb{V}\mathbb{G}(E)\mathbb{T}(E),
\end{eqnarray}
with
\begin{eqnarray}
\mathbb{V}=\left(
  \begin{array}{cc}
    V^{H_1H_2\rightarrow H_1H_2} & V^{H_1H_2\rightarrow H_3H_4} \\
    V^{H_3H_4\rightarrow H_1H_2} & V^{H_3H_4\rightarrow H_3H_4} \\
  \end{array}
\right),
\end{eqnarray}
\begin{eqnarray}
\mathbb{T}(E)=\left(
  \begin{array}{cc}
    t_{11}(E)&t_{12}(E)\\
    t_{21}(E)&t_{22}(E)\\
  \end{array}
\right),
\end{eqnarray}
and
\begin{eqnarray}
\mathbb{G}(E)=\text{diag}\{G_1(E),G_2(E)\},
\end{eqnarray}
the function $G_i$ is defined as
\begin{eqnarray}
G_i=\frac{1}{2\pi^2}\int dq\frac{q^2}{E-\sqrt{m_{i1}^2+q^2}-\sqrt{m_{i2}^2+q^2}}u^2(\Lambda).\nonumber\\
\end{eqnarray}
Here, $m_{i1}$ and $m_{i2}$ are the masses of two hadrons in the $i$-th channel, we adopt a dipole form factor $u(\Lambda)=(1+q^2/\Lambda^2)^{-2}$ \cite{Nakamura:2022gtu,Chen:2022wkh,Leinweber:2003dg} to suppress the contributions from higher momenta, we will discuss the value of cutoff $\Lambda$ in the next subsection.

\subsection{Parameters}\label{Parameters}
In Ref. \cite{Chen:2024tuu}, the parameters introduced in our model ($\tilde{g}_s$, $\tilde{g}_a$, $g_x$, and $\Lambda$) were determined from the inputs of the observed $P_{\psi}^N$ and $P_{\psi s}^\Lambda$ states. Here, we briefly review the schemes we pin down their numerical values.

Firstly, we use the experimental masses of the $P_{\psi}^N(4440)$ ($4440.0_{-5.0}^{+4.0}$ MeV) and $P_{\psi}^N(4457)$ ($4457.3_{-1.8}^{+4.0}$ MeV) \cite{LHCb:2015yax,LHCb:2019kea} as inputs to fix the $\tilde{g}_s$ and $\tilde{g}_a$ couplings. The quantum numbers of these two states have not been measured yet, thus, the assignments of these two states may have the following two scenarios
\begin{eqnarray}
\text{Scenario}\,1: P_{\psi}^N(4440)|\Sigma_c\bar{D}^*;\frac{1}{2}^-\rangle,
P_{\psi}^N(4457)|\Sigma_c\bar{D}^*;\frac{3}{2}^-\rangle,\nonumber\\\\
\text{Scenario}\,2: P_{\psi}^N(4457)|\Sigma_c\bar{D}^*;\frac{1}{2}^-\rangle,
P_{\psi}^N(4440)|\Sigma_c\bar{D}^*;\frac{3}{2}^-\rangle.\nonumber\\
\end{eqnarray}
Under these two assumptions, we performed a coupled-channel calculation by considering the $S$-wave channels composed of the $(\Lambda_c/\Sigma_c/\Sigma_c^*)+(\bar{D}/\bar{D}^*)$ hadrons \cite{Chen:2022wkh}. We fix the cutoff $\Lambda$ introduced from the dipole form factor at 1.0 GeV, and solve the coupled-channel LS equation to obtain the $\tilde{g}_s$ and $\tilde{g}_a$. In the scenario 1 and scenario 2, we find
\begin{eqnarray}
\text{Scenario}\,1&:&\tilde{g}_s=8.28\,\text{GeV}^{-2}, \,\tilde{g}_a=-1.46\,\text{GeV}^{-2},\label{scenario 1}\\
\text{Scenario}\,2&:&\tilde{g}_s=9.12\,\text{GeV}^{-2},\,\tilde{g}_a=1.25\,\text{GeV}^{-2}.\label{scenario 2}
\end{eqnarray}
We also select the other cutoff $\Lambda$ values in the ($0.5$, $1.5$) GeV region, and we can correspondingly obtain two sets of $\tilde{g}_s$ and $\tilde{g}_a$ solutions based on these two scenarios. We find that our predictions of the masses to the other $P_{\psi}^N$ states have very weak $\Lambda$-dependence in the both scenarios. Thus, we will adopt $\Lambda=1.0$ GeV throughout this work.

To estimate the uncertainties introduced from the experimental masses of $P_{\psi}^N(4440)$ and $P_{\psi}^N(4457)$ states, we combine the upper or lower limit of the mass of $P_{\psi}^N(4440)$ state with the upper or lower limit of the mass of $P_{\psi}^N(4457)$ state. In each scenario, we use these four sets of inputs to solve four sets of coupling parameters $\tilde{g}_s$ and $\tilde{g}_a$ correspondingly. We use these four sets of coupling parameters as inputs to give the lower and upper limits of other predicted states.

We further use the mass of the $P_{\psi s}^\Lambda(4338)$ to constrain the value of the SU(3) breaking factor $g_x$. Firstly, this factor is introduced to suppress the contributions from the exchanges of the isospin doublet scalar and axial-vector light mesons, the value of this factor should be smaller than 1.

Secondly, within the same framework \cite{Chen:2022wkh}, we find that in the single channel case, the mass of the $P_{\psi s}^{\Lambda}(4338)$ was obtained as 4328.1 MeV and 4323.1 MeV in the scenario 1 and scenario 2, respectively. But after including the $\Lambda_c\bar{D}_s-\Xi_{c}\bar{D}$ coupling, we find that as the $\tilde{g}_x$ increases from 0 to 0.62, the attractive force in the $\Xi_c\bar{D}$ channel becomes weak, and the mass of the $P_{\psi c}^\Lambda(4338)$ moves forward to the $\Xi_c\bar{D}$ threshold. When $g_x=0.62$, the mass of the $P_{\psi s}^\Lambda$ becomes 4335.9 MeV and 4335.7 MeV in the scenario 1 and scenario 2, respectively. When $g_x>0.62$, in the both scenarios, the attractive force between the $\Xi_c$ and $\bar{D}$ is not enough to form a bound state. Thus, due to the observation of the $P_{\psi s}^\Lambda(4338)$, we can further constrain the SU(3) breaking factor to be $g_x\leq 0.62$.

\subsection{Reduce the number of the considered di-hadron states}\label{Reduce}
The di-hadron systems listed in Table \ref{thresholds} can couple to various di-hadron states with different total isospins ($I$) and total angular momenta ($J$). We can further remove some di-hadron states that can not form bound states according to the following two rules.

Firstly, we remove the states that is forbidden according to the selection rule. For the general identical fermion and boson systems, the quantum numbers of the $H_1$ and $H_2$ hadrons satisfies the following selection rule
\begin{equation}
L+S_{\text{tot}}+I_{\text{tot}}+2i+2s=    \begin{cases}
    \text{Even}, \text{for bosons},\\
    \text{Odd},  \,\,\text{for fermions}.
    \end{cases}
\end{equation}

Secondly, we remove the states that are apparently can not form bound states. On the one hand, as can be seen from the obtained parameters $\tilde{g}_s$ and $\tilde{g}_a$ in Eqs. (\ref{scenario 1})-(\ref{scenario 2}), in both scenarios, the $\tilde{g}_s$ that is related to the flavor operator $\mathcal{O}^{\text{f}}_{\text{tot}}$ is much larger than the $\tilde{g}_a$ that is related to the flavor-spin operator $\mathcal{O}_{\text{tot}}^{\text{fs}}$. Besides, the $\tilde{g}_s$ obtained in these two scenarios have the same sign and are comparable to each other. Thus, whether the considered di-hadron state can form a bound state is flavor dominant, i.e., mainly depends on the matrix element $\langle\mathcal{O}^\text{f}_{\text{tot}}\rangle$.

On the other hand, for the di-hadron state with only one light quark in each hadron, since each hadron belongs to the flavor $\bm{3}$ representation, we have
\begin{eqnarray}
\bm{3}\otimes\bm{3}=\bm{6}\oplus\bar{\bm{3}}.
\end{eqnarray}
Thus, all the considered di-hadron states should belong to either flavor anti-symmetric $\bar{\bm{3}}$ multiplet or flavor $\bm{6}$ symmetric multiplet. From Eq. (\ref{flavorop}), the matrix elements $\langle\mathcal{O}_{\text{tot}}^{\text{f}}\rangle$ for the $\bar{\bm{3}}$ and $\bm{6}$ are obtained as $-\frac{8}{3}$ and $\frac{4}{3}$, respectively. In our convention, a negative potential and a positive effective potential will lead to an attractive force and a repulsive force, respectively. Thus, we only consider the di-hadron states that belong to the flavor $\bar{\bm{3}}$ multiplet. This is the reason that we do not list the $S=-2$ di-hadron systems in Table \ref{thresholds}, since they can not form bound states according to their symmetric flavor wave functions.

\begin{table}[htbp]
\setlength\tabcolsep{0.9pt} \caption{The remaining di-hadron states selected by the discussed two rules. We use the notations defined in Eqs. (\ref{wave1})-(\ref{wave3}) to describe the flavor-spin wave functions of the considered di-hadron states. Since all the considered di-hadron states belong to the $\bar{\bm{3}}$ flavor representation, we implicitly omit the ``$-$'' sign on the superscript of the $[H_1H_2]_J^{\frac{1}{2}-}$ state defined in Eq. (\ref{wave2}).\label{attractive states}}
\renewcommand\arraystretch{1.5}
\begin{tabular}{cccccccccc}
\toprule[1pt]
$\bar{T}_{cc}^f$&&$[\bar{D}\bar{D}^*]_{1}^0$&$[\bar{D}^*\bar{D}^*]_1^0$\\
\multirow{2}{*}{$\bar{T}^{\theta}_{cc\bar{s}}$}&&$[\bar{D}\bar{D}_s^*]_{1}$&\multirow{2}{*}{$[\bar{D}^*\bar{D}_s^*]_{1}^{\frac{1}{2}}$}\\
                          &&$[\bar{D}_s\bar{D}^*]_{1}$\\
\hline
$P_{\psi c}^{\Lambda}$&$[\Xi_{cc}\bar{D}]_{\frac{1}{2}}^{0}$&$[\Xi^*_{cc}\bar{D}]_{\frac{3}{2}}^{0}$&$[\Xi_{cc}\bar{D}^*]^{0}_{\frac{1}{2},\frac{3}{2}}$&$[\Xi_{cc}^*\bar{D}^*]^{0}_{\frac{1}{2},\frac{3}{2},\frac{5}{2}}$\\
\multirow{2}{*}{$P_{\psi cs}^N$}&$[\Xi_{cc}\bar{D}_s]_{\frac{1}{2}}$&$[\Xi_{cc}^*\bar{D}_s]_{\frac{3}{2}}$&$[\Xi_{cc}\bar{D}_s^*]_{\frac{1}{2},\frac{3}{2}}$&$[\Xi_{cc}^*\bar{D}_s^*]_{\frac{1}{2},\frac{3}{2},\frac{5}{2}}$\\
&$[\Omega_{cc}\bar{D}]_{\frac{1}{2}}$&$[\Omega_{cc}^*\bar{D}]_{\frac{3}{2}}$&$[\Omega_{cc}\bar{D}^*]_{\frac{1}{2},\frac{3}{2}}$&$[\Omega_{cc}^*\bar{D}^*]_{\frac{1}{2},\frac{3}{2},\frac{5}{2}}$\\
\hline
$H_{\Omega_{ccc}c}^{\Lambda}$&$[\Xi_{cc}\Xi_{cc}]_1^0$&$[\Xi_{cc}\Xi_{cc}^*]_1^0$&$[\Xi_{cc}\Xi_{cc}^*]_2^0$&$[\Xi_{cc}^*\Xi_{cc}^*]_{1,3}^0$\\
\multirow{2}{*}{$H_{\Omega_{ccc}cs}^{N}$}&$[\Xi_{cc}\Omega_{cc}]^{\frac{1}{2}}_1$&$[\Xi_{cc}\Omega^*_{cc}]_1$&$[\Xi_{cc}\Omega^*_{cc}]_2$&$[\Xi_{cc}^*\Omega_{cc}^*]^{\frac{1}{2}}_{0,3}$\\
&&$[\Xi_{cc}^*\Omega_{cc}]_1$&$[\Xi_{cc}^*\Omega_{cc}]_2$\\
\bottomrule[1pt]
\end{tabular}
\end{table}
According to the above two rules, We remove some of the di-hadron states composed of the systems listed in Table \ref{thresholds}, then we collect the rest of the di-hadron states and present them in Table \ref{attractive states}. Note that in Table \ref{attractive states}, we adopt the wave function notations defined in Eqs. (\ref{wave1})-(\ref{wave3}). Especially, we implicitly omit the sign on the superscript of the wave function $[H_1H_2]_{J}^{\frac{1}{2}(\pm)}$ since we only consider the states that belong to the $\bar{\text{3}}$ flavor multiplet. In the following, we will only investigate the states listed in Table \ref{attractive states}.
\section{Numerical results}\label{Numerical}
We calculate the matrix elements of the $\bar{T}_{cc}^f$/$\bar{T}_{cc\bar{s}}^\theta$, $P_{\psi c}^\Lambda$/$P_{\psi cs}^N$, and $H_{\Omega_{ccc}c}^\Lambda$/$H_{\Omega_{ccc}cs}^N$ states collected in Table \ref{attractive states} and present the obtained [$\langle\mathcal{O}_{\text{tot}}^\text{f}\rangle$, $\langle\mathcal{O}_{\text{tot}}^\text{fs}\rangle$] in Table \ref{matrixele}.

\begin{table*}[htbp]
\setlength\tabcolsep{1pt} \caption{The matrix elements [$\langle\mathcal{O}^{\text{f}}_{\text{tot}}\rangle$,$\langle\mathcal{O}^{\text{fs}}_{\text{tot}}\rangle$] of the $\bar{T}_{cc}^f$/$\bar{T}_{cc\bar{s}}^\theta$, $P_{\psi c}^\Lambda$/$P_{\psi cs}^N$, and $H_{\Omega_{ccc}c}^\Lambda$/$H_{\Omega_{ccc}cs}^N$ states collected in Table \ref{attractive states}. We consider the isospin breaking for the $\bar{T}_{cc}^f$/$\bar{T}_{cc\bar{s}}^\theta$ states, and adopt the isospin limit but consider the SU(3) breaking for the $P_{\psi c}^\Lambda$/$P_{\psi cs}^N$ and $H_{\Omega_{ccc}c}^\Lambda$/$H_{\Omega_{ccc}cs}^N$ states.\label{matrixele}}
\renewcommand\arraystretch{1.5}
\begin{tabular}{cccccccccc}
\hline
$\bar{T}_{cc}^f$&$[\langle\mathcal{O}^{\text{f}}_{\text{tot}}\rangle,\langle\mathcal{O}^{\text{fs}}_{\text{tot}}\rangle]$&$\bar{T}^{\theta}_{cc\bar{s}}$&$[\langle\mathcal{O}^{\text{f}}_{\text{tot}}\rangle,\langle\mathcal{O}^{\text{fs}}_{\text{tot}}\rangle]$&$\langle\mathcal{O}^{\text{f}}_{\text{Eig}}\rangle$&$\langle\mathcal{O}^{\text{fs}}_{\text{Eig}}\rangle$\\
($[\bar{D}^0D^{*-}]_1,$&\multirow{2}{*}{$\left(
\begin{array}{cc}
[-\frac{2}{3},0] & [2,0] \\
& [-\frac{2}{3},0] \\
\end{array}
\right)$}&$([\bar{D}^0D_s^{*-}]_1,$&\multirow{2}{*}{$\left(
\begin{array}{cc}
[-\frac{2}{3},0] & [2g_x,0] \\
& [-\frac{2}{3},0] \\
\end{array}
\right)$}&\multirow{2}{*}{$\left(
\begin{array}{c}
-\frac{8}{3} \\
 \frac{4}{3}\\
\end{array}
\right)$}&\multirow{2}{*}{$\left(
\begin{array}{c}
 0\\
 0\\
\end{array}
\right)$}\\
$[D^-\bar{D}^{*0}]_1)$&&$[D_s^-\bar{D}^{*0}]_1)$\\
$[\bar{D}^{*0}D^{*-}]^0_1$&$[-\frac{8}{3},\frac{8}{3}]$
&$[\bar{D}^{*0}D_s^{*-}]^{\frac{1}{2}}_1$&$[-2g_x-\frac{2}{3},2g_x+\frac{2}{3}]$&\multirow{1}{*}{$\left(
\begin{array}{c}
-\frac{8}{3} \\
\end{array}
\right)$}&\multirow{1}{*}{$\left(
\begin{array}{c}
\frac{8}{3} \\
\end{array}
\right)$}\\
\hline
$P^{\Lambda}_{\psi c}$&$[\langle\mathcal{O}_{\text{tot}}^{\text{f}}\rangle,\langle\mathcal{O}_{\text{tot}}^{\text{fs}}\rangle]$&$P^N_{\psi cs}$&$[\langle\mathcal{O}^{\text{f}}_{\text{tot}}\rangle,\langle\mathcal{O}^{\text{fs}}_{\text{tot}}\rangle]$&$\langle\mathcal{O}_{\text{Eig}}^{\text{f}}\rangle$&$\langle\mathcal{O}_{\text{Eig}}^{\text{fs}}\rangle$\\
\hline
\multirow{2}{*}{$[\Xi_{cc}\bar{D}]_{\frac{1}{2}}^0$}&\multirow{2}{*}{$[-\frac{8}{3},0]$}&$([\Xi_{cc}\bar{D}_s]_{\frac{1}{2}},$&\multirow{2}{*}{$\left(
\begin{array}{cc}
[-\frac{2}{3},0] & [2g_x,0] \\
& [-\frac{2}{3},0] \\
\end{array}
\right)$}&\multirow{1}{*}{$\left(
\begin{array}{c}
-\frac{8}{3} \\
 \frac{4}{3}\\
\end{array}
\right)$}&\multirow{1}{*}{$\left(
\begin{array}{c}
0 \\
0\\
\end{array}
\right)$}&
\\
                                &&$[\Omega_{cc}\bar{D}]_{\frac{1}{2}})$&&&\\
\multirow{2}{*}{$[\Xi_{cc}\bar{D}^*]^0_{\frac{1}{2}}$}&\multirow{2}{*}{$[-\frac{8}{3},-\frac{16}{9}]$}                                &$([\Xi_{cc}\bar{D}_s^{*}]_{\frac{1}{2}},$&\multirow{2}{*}{$\left(
\begin{array}{cc}
[-\frac{2}{3},-\frac{4}{9}] & [2g_x,\frac{4}{3}g_x] \\
& [-\frac{2}{3},-\frac{4}{9}] \\
\end{array}
\right)$}&\multirow{1}{*}{$\left(
\begin{array}{c}
-\frac{8}{3} \\
\frac{4}{3}\\
\end{array}
\right)$}&\multirow{1}{*}{$\left(
\begin{array}{c}
-\frac{16}{9} \\
\frac{8}{9}\\
\end{array}
\right)$}
&\\
&&$[\Omega_{cc}\bar{D}^{*}]_{\frac{1}{2}})$&&&\\
\multirow{2}{*}{$[\Xi_{cc}^*\bar{D}^*]_{\frac{1}{2}}^0$}&\multirow{2}{*}{$[-\frac{8}{3},\frac{40}{9}]$}                                &$([\Xi_{cc}^{*}\bar{D}_s^{*}]_{\frac{1}{2}},$&\multirow{2}{*}{$\left(
\begin{array}{cc}
[-\frac{2}{3},\frac{10}{9}] & [2g_x,-\frac{10}{3}g_x] \\
& [-\frac{2}{3},\frac{10}{9}] \\
\end{array}
\right)$}&\multirow{1}{*}{$\left(
\begin{array}{c}
 -\frac{8}{3}\\
\frac{4}{3}\\
\end{array}
\right)$}&\multirow{1}{*}{$\left(
\begin{array}{c}
\frac{40}{9} \\
-\frac{20}{9}\\
\end{array}
\right)$}&
&\\
&                                &$[\Omega_{cc}^{*}\bar{D}^{*}]_{\frac{1}{2}})$&&&\\
\hline
\multirow{2}{*}{$[\Xi_{cc}^*\bar{D}]^0_{\frac{3}{2}}$}&\multirow{2}{*}{$[-\frac{8}{3},0]$}
&$([\Xi_{cc}^{*}\bar{D}_s]_{\frac{3}{2}},$&\multirow{2}{*}{$\left(
\begin{array}{cc}
[-\frac{2}{3},0] & [2g_x,0] \\
& [-\frac{2}{3},0] \\
\end{array}
\right)$}&\multirow{1}{*}{$\left(
\begin{array}{c}
-\frac{8}{3} \\
\frac{4}{3}\\
\end{array}
\right)$}&\multirow{1}{*}{$\left(
\begin{array}{c}
0\\
0\\
\end{array}
\right)$}&
\\
&                     &$[\Omega_{cc}^{*}\bar{D}]_{\frac{3}{2}})$&&&\\
\multirow{2}{*}{$[\Xi_{cc}\bar{D}^*]_{\frac{3}{2}}^0$}&\multirow{2}{*}{$[-\frac{8}{3},\frac{8}{9}]$}                                &$([\Xi_{cc}\bar{D}_s^{*}]_{\frac{3}{2}},$&\multirow{2}{*}{$\left(
\begin{array}{cc}
[-\frac{2}{3},\frac{2}{9}] & [2g_x,-\frac{2}{3}g_x] \\
& [-\frac{2}{3},\frac{2}{9}] \\
\end{array}
\right)$}&\multirow{1}{*}{$\left(
\begin{array}{c}
-\frac{8}{3} \\
\frac{4}{3}\\
\end{array}
\right)$}&\multirow{1}{*}{$\left(
\begin{array}{c}
\frac{8}{9} \\
-\frac{4}{9}\\
\end{array}
\right)$}
&\\
&                                &$[\Omega_{cc}\bar{D}^{*}]_{\frac{3}{2}})$&&&\\
\multirow{2}{*}{$[\Xi_{cc}^*\bar{D}^*]_{\frac{3}{2}}^0$}&\multirow{2}{*}{$[-\frac{8}{3},\frac{16}{9}]$}                                &$([\Xi_{cc}^{*}\bar{D}_s^{*}]_{\frac{3}{2}},$&\multirow{2}{*}{$\left(
\begin{array}{cc}
[-\frac{2}{3},\frac{4}{9}] & [2g_x,-\frac{4}{3}g_x] \\
& [-\frac{2}{3},\frac{4}{9}] \\
\end{array}
\right)$}&\multirow{1}{*}{$\left(
\begin{array}{c}
-\frac{8}{3} \\
\frac{4}{3}\\
\end{array}
\right)$}&\multirow{1}{*}{$\left(
\begin{array}{c}
\frac{16}{9} \\
-\frac{8}{9}\\
\end{array}
\right)$}
&&\\
&                                &$[\Omega_{cc}^{*}\bar{D}^{*}]_{\frac{3}{2}})$&&&\\
\hline
\multirow{2}{*}{$[\Xi_{cc}^*\bar{D}^*]_{\frac{5}{2}}^0$}&\multirow{2}{*}{$[-\frac{8}{3},-\frac{8}{3}]$}&$([\Xi_{cc}^{*}\bar{D}_s^{*}]_{\frac{5}{2}},$&\multirow{2}{*}{$\left(
\begin{array}{cc}
[-\frac{2}{3},-\frac{2}{3}] & [2g_x,2g_x] \\
& [-\frac{2}{3},-\frac{2}{3}] \\
\end{array}
\right)$}&\multirow{1}{*}{$\left(
\begin{array}{c}
-\frac{8}{3} \\
\frac{4}{3}\\
\end{array}
\right)$}&\multirow{1}{*}{$\left(
\begin{array}{c}
-\frac{8}{3} \\
\frac{4}{3}\\
\end{array}
\right)$}
&\\
&                                &$[\Omega_{cc}^{*}\bar{D}^{*}]_{\frac{5}{2}})$&&&\\
\hline
$H^{\Lambda}_{\Omega_{ccc}c}$&$[\langle\mathcal{O}^{\text{f}}_{\text{tot}}\rangle,\langle\mathcal{O}^{\text{fs}}_{\text{tot}}\rangle]$&$H^{N}_{\Omega_{ccc}cs}$&$[\langle\mathcal{O}^{\text{f}}_{\text{tot}}\rangle,\langle\mathcal{O}^{\text{fs}}_{\text{tot}}\rangle]$&$\langle\mathcal{O}^{\text{f}}_{\text{Eig}}\rangle$&$\langle\mathcal{O}^{\text{fs}}_{\text{Eig}}\rangle$\\
\hline
$[\Xi_{cc}\Xi_{cc}]_1^0$&$[-\frac{8}{3},-\frac{8}{27}]$
&$[\Xi_{cc}\Omega_{cc}]_1^{\frac{1}{2}}$&$[-2g_x-\frac{2}{3},-\frac{2}{9}g_x-\frac{2}{27}]$&\multirow{1}{*}{$\left(
\begin{array}{c}
-\frac{8}{3} \\
\end{array}
\right)$}&\multirow{1}{*}{$\left(
\begin{array}{c}
-\frac{8}{27} \\
\end{array}
\right)$}\\
\multirow{2}{*}{$[\Xi_{cc}\Xi_{cc}^*]_1^0$}&\multirow{2}{*}{$[-\frac{8}{3},-\frac{40}{27}]$}
&($[\Xi_{cc}\Omega_{cc}^*]_1,$&\multirow{2}{*}{$\left(
\begin{array}{cc}
[-\frac{2}{3},-\frac{10}{27}] & [2g_x,\frac{10}{9}g_x] \\
& [-\frac{2}{3},-\frac{10}{27}] \\
\end{array}
\right)$}&\multirow{2}{*}{$\left(
\begin{array}{c}
-\frac{8}{3} \\
\frac{4}{3}\\
\end{array}
\right)$}&\multirow{2}{*}{$\left(
\begin{array}{c}
-\frac{40}{27} \\
\frac{20}{27}\\
\end{array}
\right)$}\\
&&$[\Omega_{cc}\Xi^*_{cc}]_1)$&\\
$[\Xi^*_{cc}\Xi_{cc}^*]_1^0$&$[-\frac{8}{3},\frac{88}{27}]$
&$[\Xi_{cc}^*\Omega_{cc}^*]_1^{\frac{1}{2}}$&$[-2g_x-\frac{2}{3},\frac{22}{9}g_x+\frac{22}{27}]$&\multirow{1}{*}{$\left(
\begin{array}{c}
-\frac{8}{3} \\
\end{array}
\right)$}&\multirow{1}{*}{$\left(
\begin{array}{c}
\frac{88}{27} \\
\end{array}
\right)$}\\
\hline
\multirow{2}{*}{$[\Xi_{cc}\Xi_{cc}^*]_2^{0}$}&\multirow{2}{*}{$[-\frac{8}{3},\frac{8}{9}]$}
&($[\Xi_{cc}\Omega^*_{cc}]_2,$&\multirow{2}{*}{$\left(
\begin{array}{cc}
[-\frac{2}{3},\frac{2}{9}] & [2g_x,-\frac{2}{3}g_x] \\
& [-\frac{2}{3},\frac{2}{9}] \\
\end{array}
\right)$}&\multirow{2}{*}{$\left(
\begin{array}{c}
-\frac{8}{3} \\
 \frac{4}{3}\\
\end{array}
\right)$}&\multirow{2}{*}{$\left(
\begin{array}{cc}
\frac{8}{9} \\
-\frac{4}{9} \\
\end{array}
\right)$}\\
&&$[\Omega_{cc}\Xi^*_{cc}]_2)$&\\
\hline
$[\Xi_{cc}^*\Xi_{cc}^*]_3^0$&$[-\frac{8}{3},-\frac{8}{3}]$&$[\Xi_{cc}^*\Omega_{cc}^*]_3^{\frac{1}{2}}$&$[-2g_x-\frac{2}{3},-2g_x-\frac{2}{3}]$
&\multirow{1}{*}{$\left(
\begin{array}{c}
-\frac{8}{3} \\
\end{array}
\right)$}&\multirow{1}{*}{$\left(
\begin{array}{c}
-\frac{8}{3} \\
\end{array}
\right)$}\\
\bottomrule[1pt]
\end{tabular}
\end{table*}

As presented in Table \ref{matrixele}, since the mass of $\bar{T}_{cc}^f(3875)$ is very close to the $\bar{D}^0D^{*-}$ and $D^-\bar{D}^{*0}$ thresholds, thus, we take into account mass difference between the $\bar{D}^0D^{*-}$ and $D^-\bar{D}^{*0}$ thresholds and mix these two components through a 2$\times$2 matrix. We can further obtain the strangeness partner of the $\bar{T}_{cc}^f$ states by replacing a $d$ quark to an $s$ quark. Note that the effective potentials of the $\bar{T}_{cc\bar{s}}^\theta$ states are slightly different from that of the $\bar{T}_{cc}^f$ states by a SU(3) breaking factor $g_x$, in the SU(3) limit with $g_x=1$, the $\bar{T}_{cc}^f$ states and their strangeness partner $\bar{T}_{cc\bar{s}}^\theta$ states share identical effective potentials.

The eigenvalues of the $\bar{T}_{cc}^f$ matrices calculated from the operators $\mathcal{O}_{\text{tot}}^\text{f}$ and $\mathcal{O}_{\text{tot}}^\text{fs}$ are also presented in the fifth and sixth columns of Table \ref{matrixele}, respectively. Besides, one can easily find that in the SU(3) limit with $g_x=1$, the $\bar{T}_{cc}^f$ and its strangeness partner $\bar{T}_{ccs}^{\theta}$ have identical effective potential matrices, and thus have identical $\langle\mathcal{O}_{\text{Eig}}^\text{f}\rangle$ and $\langle\mathcal{O}_{\text{Eig}}^\text{fs}\rangle$ matrices.

For the $P_{\psi c}^{\Lambda}$ states, we adopt the isospin limit and calculate the effective potentials of the $\Xi_{cc}^{(*)}\bar{D}^{(*)}$ states in the single-channel formalism. However, as demonstrated in Table \ref{matrixele}, for the $P_{\psi cs}^N$ states, the strange quark belongs to the baryon or meson will lead to different final states, this is another source of the SU(3) breaking. As presented in table \ref{matrixele}, we include this effect by distinguishing these two components and calculating the effective potentials of the $P_{\psi cs}^N$ states through 2$\times$2 matrices. In the SU(3) limit with $g_x=1$, the mixture of the $\Xi_{cc}^{(*)}\bar{D}_s^{(*)}$ and $\Omega_{cc}^{(*)}\bar{D}^{(*)}$ components will lead to two different flavor matrix elements $-\frac{8}{3}$ and $\frac{4}{3}$, they belong to the $\bar{\bm{3}}$ and $\bm{6}$ flavor multiplets, respectively. As discussed in Sec. \ref{Reduce}, only the $P_{\psi cs}^N$ state with flavor matrix element $-\frac{8}{3}$ has attractive force and thus may form bound state.

The effective potentials of the $H_{\Omega_{ccc}c}^\Lambda$/$H_{\Omega_{ccc}cs}^N$ states are also presented in the lower panel of Table \ref{matrixele}. Similar to the $P_{\psi c}^\Lambda$ states, for the $H_{\Omega_{ccc}c}^\Lambda$ states, we adopt the isospin limit to calculate the effective potentials of the $\Xi_{cc}^{(*)}\Xi_{cc}^{(*)}$ states. But for the $H_{\Omega_{ccc}cs}^N$ states, the $\Xi_{cc}\Omega_{cc}^*$ and $\Omega_{cc}\Xi_{cc}^*$ components have different thresholds, we mix these two components to include the SU(3) breaking effect induced from the constituent strange quark mass.

With the effective potentials for all the considered $\bar{T}_{cc}^{f}$/$\bar{T}_{cc\bar{s}}^{\theta}$/$P_{\psi c}^\Lambda$/$P_{\psi cs}^N$/$H_{\Omega_{ccc}c}^\Lambda$/$H_{\Omega_{ccc}cs}^N$ states listed in Table \ref{matrixele}, we find the possible bound states via solving the corresponding LS equations. Here, with the parameters from the scenario 1 and scenario 2 in Eqs. (\ref{scenario 1})-(\ref{scenario 2}), we firstly discuss the possible bound states and their spectra among the $\bar{T}_{cc}^f$/$P_{\psi c}^\Lambda$/$H_{\Omega_{ccc}c}^\Lambda$ states. Then we discuss the existence of the bound states in the $\bar{T}_{cc\bar{s}}^{\theta}$/$P_{\psi cs}^N$/$H_{\Omega_{ccc}cs}^N$ systems by including the SU(3) breaking effects. We proceed to discuss their possible mass spectra if they do exist.

Note that in this work, we use the exactly identical theoretical framework and parameters to that of our previous work \cite{Chen:2024tuu}, where we present a detailed investigation on the flavor-spin symmetry properties of the interactions among the $P_{\psi}^N$/$P_{\psi s}^\Lambda$/$H_{\Omega_{ccc}}^N$/$H_{\Omega_{ccc}s}^\Lambda$ systems. Then the flavor-spin symmetry allow us to relate the interactions of the $\bar{T}_{cc}^f$/$\bar{T}_{cc\bar{s}}^\theta$/$P_{\psi c}^\Lambda$/$P_{\psi cs}^N$/$H_{\Omega_{ccc}c}^\Lambda$/$H_{\Omega_{ccc}cs}^N$ systems studied in this work to the $P_{\psi}^N$/$P_{\psi s}^\Lambda$/$H_{\Omega_{ccc}}^N$/$H_{\Omega_{ccc}s}^\Lambda$ systems discussed in Ref. \cite{Chen:2024tuu}.

\subsection{Non-strange $\bar{T}_{cc}^f$, $P_{\psi c}^\Lambda$, and $H_{\Omega_{ccc}c}^\Lambda$ states}\label{nonstrange}
With the parameters $\tilde{g}_s$ and $\tilde{g}_a$ from the scenario 1 and scenario 2 listed in Eqs. (\ref{scenario 1})-(\ref{scenario 2}), we search for the bound state solutions from the considered di-hadron $\bar{T}_{cc}^f$/$P_{\psi c}^\Lambda$/$H_{\Omega_{ccc}c}^\Lambda$ states, and collect their binding energies in Table \ref{binding energy}. We use the experimental uncertainties from the masses of the $P_{\psi}^N(4440)$ and $P_{\psi}^N(4457)$ states to give the lower and upper limits of our predictions. In Table \ref{binding energy}, the ``$+*$'' denotes that the bound state solution does not exist at its upper limit.

\begin{table}[htbp]
\setlength\tabcolsep{0.9pt} \caption{The binding energies of the non-strange $\bar{T}_{cc}^f$, $P_{\psi c}^\Lambda$, and $H_{\Omega_{ccc}c}^\Lambda$ bound states. The ``*'' denotes that the bound state solution no longer exists at its upper limit. All the results are in units of MeV. \label{binding energy}}
\renewcommand\arraystretch{1.5}
\begin{tabular}{cccccccccc}
\toprule[1pt]
&\multicolumn{2}{c}{Scenario 1}&\multicolumn{2}{c}{Scenario 2}\\
System&Mass&BE&Mass&BE\\
\hline
$[\bar{D}\bar{D}^*]_1^0$&$3875.0^{+*}_{-0.3}$&$-0.1_{-0.3}^{+*}$&$3874.1^{+1.0}_{-1.2}$&$-1.0^{+1.0}_{-1.2}$\\
$[\bar{D}^*\bar{D}^*]_1^0$&$4013.1_{-2.6}^{+1.4}$&$-4.1_{-2.6}^{+1.4}$&$4016.9^{+*}_{-0.5}$&$-0.2^{+*}_{-0.5}$\\
\hline
$[\Xi_{cc}\bar{D}]_{\frac{1}{2}}^0$&$5484.2^{+2.6}_{-1.8}$&$-4.4_{-1.8}^{+2.6}$&$5480.8_{-2.5}^{+2.9}$&$-7.9_{-2.5}^{+2.9}$\\
$[\Xi_{cc}\bar{D}^*]_{\frac{1}{2}}^0$&$5627.7_{-1.4}^{+1.8}$&$-2.3_{-1.4}^{+1.8}$&$5616.1_{-3.1}^{+3.7}$&$-13.9_{-3.1}^{+3.7}$\\
$[\Xi_{cc}^*\bar{D}^*]_{\frac{1}{2}}^0$&$5716.7_{-3.0}^{+5.1}$&$-18.8_{-3.0}^{+5.1}$&$5733.8_{-1.5}^{+1.3}$&$-1.8_{-1.5}^{+1.3}$\\
$[\Xi_{cc}^*\bar{D}]_{\frac{3}{2}}^0$&$5589.6_{-1.8}^{+2.7}$&$-4.7_{-1.8}^{+2.7}$&$5586.0_{-2.6}^{+3.0}$&$-8.2_{-2.6}^{+3.0}$\\
$[\Xi_{cc}\bar{D}^*]_{\frac{3}{2}}$&$5622.2_{-2.4}^{+3.1}$&$-7.7_{-2.4}^{+3.1}$&$5622.5_{-2.3}^{+3.2}$&$-7.5_{-2.3}^{+3.2}$\\
$[\Xi_{cc}^*\bar{D}^*]_{\frac{3}{2}}$&$5725.1_{-3.0}^{+2.9}$&$-10.5_{-3.0}^{+2.9}$&$5729.6_{-2.0}^{+3.1}$&$-6.0_{-2.0}^{+3.1}$\\
$[\Xi_{cc}^*\bar{D}^*]_{\frac{5}{2}}$&$5734.3_{-1.5}^{+*}$&$-1.2_{-1.5}^{+*}$&$5718.8_{-4.5}^{+3.6}$&$-16.7_{-4.5}^{+3.6}$\\
\hline
$[\Xi_{cc}\Xi_{cc}]_1^0$&$7227.0_{-3.0}^{+5.5}$&$-15.8_{-3.0}^{+5.5}$&$7219.0_{-4.1}^{+4.8}$&$-23.8_{-4.1}^{+4.8}$\\
$[\Xi_{cc}\Xi_{cc}^*]_1^0$&$7336.2_{-2.4}^{+5.8}$&$-12.2_{-2.4}^{+5.8}$&$7319.7_{-4.7}^{+4.4}$&$-28.7_{-4.7}^{+4.4}$\\
$[\Xi_{cc}^*\Xi_{cc}^*]_1^0$&$7422.7_{-7.9}^{+4.6}$&$-31.3_{-7.9}^{+4.6}$&$7440.5_{-2.5}^{+5.5}$&$-13.5_{-2.5}^{+5.5}$\\
$[\Xi_{cc}\Xi_{cc}^*]_2^0$&$7327.6_{-3.7}^{+5.0}$&$-20.8_{-3.7}^{+5.0}$&$7328.0_{-3.6}^{+5.2}$&$-20.4_{-3.6}^{+5.2}$\\
$[\Xi_{cc}^*\Xi_{cc}^*]_3^0$&$7445.1_{-3.2}^{+5.7}$&$-8.9_{-3.2}^{+5.7}$&$7420.1_{-6.0}^{+4.9}$&$-33.9_{-6.0}^{+4.9}$\\
\bottomrule[1pt]
\end{tabular}
\end{table}

As presented in Table \ref{matrixele}, for the $[\bar{D}\bar{D}^*]_1^0$ state, we consider the isospin breaking effect by including the mixture of the $[\bar{D}^0D^{*-}]_1$ and $[D^-\bar{D}^{*0}]_1$ components. The threshold of the $\bar{D}^0D^{*-}$ (3875.1 MeV) is slightly below the threshold of the $D^-\bar{D}^{*0}$ (3876.5 MeV). We search for the bound state solution below the $\bar{D}^0D^{*-}$ threshold and we also check the quasi-bound state solution in the energy region $E_{\bar{D}^0D^{*-}}<E<E_{D^{-}\bar{D}^{*0}}$. We can only find one bound state which is below the lowest $\bar{D}^0D^{*-}$ channel.

The obtained binding energies for the $[\bar{D}\bar{D}^*]_1^0$ and $[\bar{D}^*\bar{D}^*]_1^0$ states in these two scenarios are very interesting. As presented in Table \ref{binding energy}, in the scenario 1, the absolute value of the binding energy for the $[\bar{D}^*\bar{D}^*]_1^0$ state is larger than that of the $[\bar{D}\bar{D}^*]_1^0$ state. This result suggests that if the $J^P$ of the observed $P_{\psi}^N(4440)$ and $P_{\psi}^N(4457)$ states were measured to be $1/2^-$ and $3/2^-$, respectively, then this measurement would give a strong support to the existence of a $[\bar{D}^*\bar{D}^*]_1^0$ state with its absolute value of binding energy larger than that of the observed $T_{cc}^f(3875)$ state.

Similarly, as presented in Table \ref{binding energy}, in the scenario 2, we find that the absolute value of the binding energy for the $[\bar{D}^*\bar{D}^*]_1^0$ state is smaller than that of the $[\bar{D}\bar{D}^*]_1^0$ state. This result suggests that if the $J^P$ of the observed $P_{\psi}^N(4440)$ and $P_{\psi}^N(4457)$ were measured to be $3/2^-$ and $1/2^-$, respectively, then this measurement would suggest that the $[\bar{D}^*\bar{D}^*]_1^0$ should have a very minor binding energy or could not exist.

Or vise versa, the observation/non-observation of the $[\bar{D}^*\bar{D}^*]_1^0$ bound state and the measurement of its binding energy would give strong hints to the $J^P$ quantum numbers of the $P_{\psi}^N(4440)$ and $P_{\psi}^N(4457)$ states. Then the $J^P$ quantum numbers of the $P_{\psi}^N(4440)$ and $P_{\psi}^N(4457)$ states can be further related to the $J^P$ quantum numbers of the other predicted $P_{\psi}^N$ states according to the flavor-spin symmetry discussed in our previous work \cite{Chen:2024tuu}.

The binding energies of the $P_{\psi c}^{\Lambda}$ and $H_{\Omega_{ccc}c}^{\Lambda}$ calculated from the scenario 1 and scenario 2 are also presented in Table \ref{binding energy} (See also the calculations in Refs.~\cite{Wang:2023eng,Wang:2024riu}). As listed in Table \ref{matrixele}, the effective potentials of the $[\Xi_{cc}^*\bar{D}^*]_{\frac{5}{2}}^0$ and $[\Xi_{cc}^*\Xi_{cc}^*]_3^0$ are the same, thus, the binding energies given in Table \ref{binding energy} show that with the same effective potential, the state with larger reduced mass binds deeper.

\begin{figure*}[htbp]
    \centering
    \includegraphics[width=1.0\linewidth]{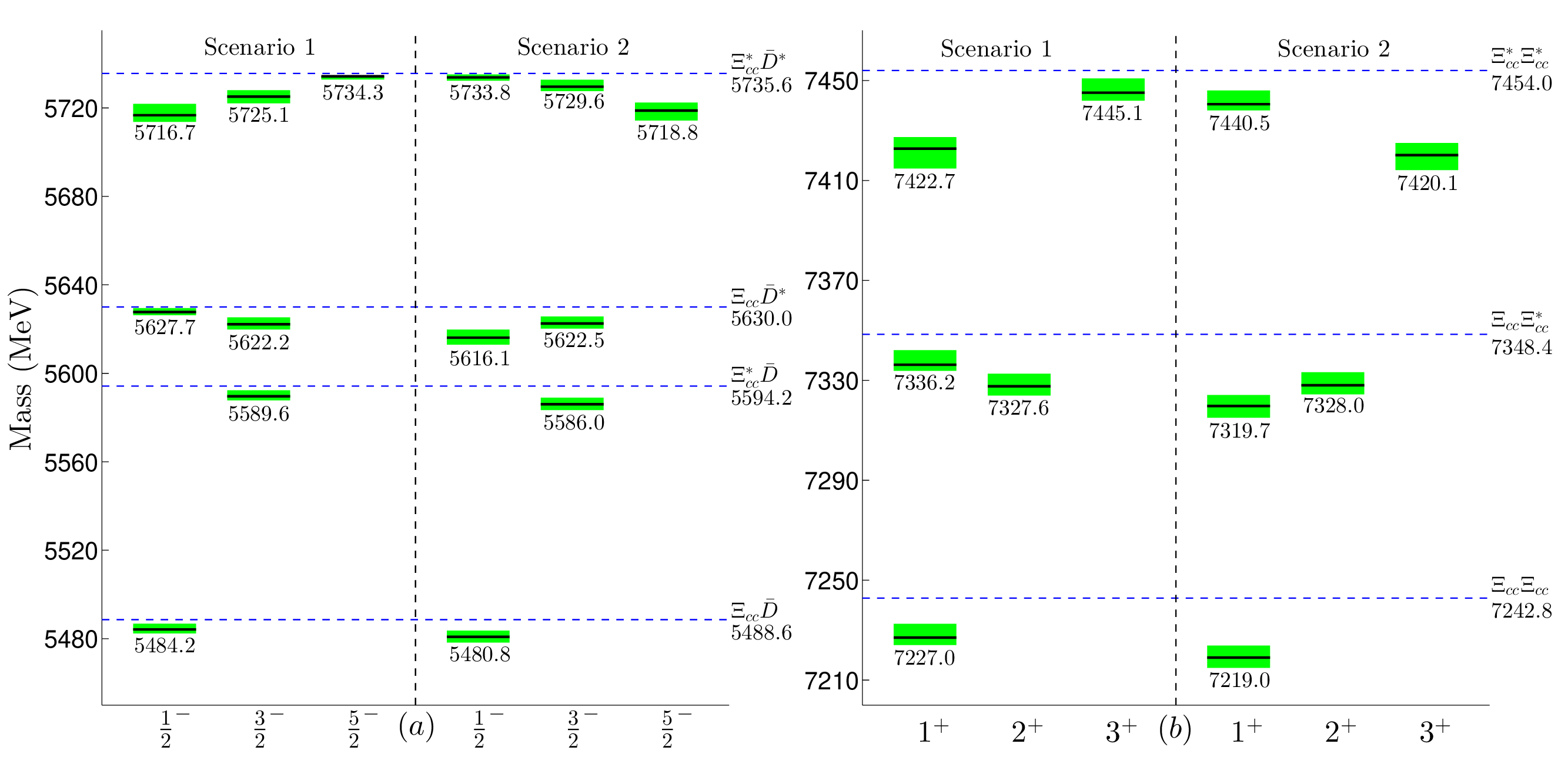}
    \caption{The mass spectra of the $P_{\psi c}^\Lambda$ and $H_{\Omega_{ccc}c}^{\Lambda}$ bound states. The results of the $P_{\psi c}^{\Lambda}$ bound states for the scenario 1 and scenario 2 are illustrated on the left and right sides of (a), respectively. The results of the $H_{\Omega_{ccc}c}^\Lambda$ bound states for the scenario 1 and scenario 2 are illustrated on the left and right sides of (b), respectively. We plot the central values of the $P_{\psi c}^\Lambda$ and $H_{\Omega_{ccc}c}^{\Lambda}$ bound states with black lines and label the corresponding masses. The theoretical errors are introduced by considering the experimental errors from the masses of the $P_{\psi}^N(4440)$ and $P_{\psi}^N(4457)$ states, they are labeled with green bands.}
    \label{massspectra}
\end{figure*}

We plot the mass spectra of the $P_{\psi c}^N$ bound states obtained from the scenario 1 and scenario 2 on the left and right sides of Fig. \ref{massspectra} (a), respectively. Note that the flavor-spin coupling $\tilde{g}_a$ determined in the scenario 1 and scenario 2 have opposite signs, thus, the scenario 1 and scenario 2 give different mass arrangements to the mass spectra of the $P_{\psi c}^N$ bound states. As can be checked from Fig. \ref{massspectra} (a), in the scenario 1, the mass of the bound state that is composed of the $\Xi_{cc}\bar{D}^*$ decreases as the total momentum $J$ increases, while the mass of the bound state that is composed of the $\Xi_{cc}^*\bar{D}^*$ increases as the total angular momentum $J$ increases. These tendencies are opposite in the scenario 2 case.

We also plot the mass spectra of the $H_{\Omega_{ccc}c}^{\Lambda}$ bound states obtained from the scenario 1 and scenario 2 on the left and right sides of Fig. \ref{massspectra} (b), respectively. As can be checked from Fig. \ref{massspectra} (b), similar to the mass spectra of the $P_{\psi c}^N$ states, different scenarios also lead to different mass arrangements of the $\Xi_{cc}\Xi_{cc}^*$ or $\Xi_{cc}^*\Xi^*_{cc}$ bound states with different total angular momenta $J$.
\subsection{The existences of the bound states in the $\bar{T}_{cc\bar{s}}^\theta$/$P_{\psi cs}^N$/$H_{\Omega_{ccc}cs}^N$ systems}
In Sec. \ref{nonstrange}, with the parameters extracted from the inputs of observed $P_{\psi}^N(4440)$ and $P_{\psi}^N(4457)$ states, we successfully reproduce the mass of observed $T^f_{cc}(3875)$ state, indicating the similar binding mechanism for the $P_{\psi}^N$ and $T_{cc}^f$ bound states. Then we proceed to use the parameters extracted from the inputs of observed $P_{\psi}^N(4440)$, $P_{\psi}^N(4457)$, and $P_{\psi s}^\Lambda(4338)$ states to discuss the existences of the bound states in the $\bar{T}_{cc\bar{s}}^\theta$/$P_{\psi cs}^N$/$H_{\Omega_{ccc}cs}^N$ systems.

Since the observed $\bar{T}_{cc}^f(3875)$ has a very tiny binding energy, implying that the interaction between the $\bar{D}$ and $\bar{D}^*$ is just enough to bind them together. Thus, to clarify the existences of its strangeness partner $\bar{T}_{cc\bar{s}}^\theta$ as well as its HDAS partners $P_{\psi cs}^N$ and $H_{\Omega_{ccc}cs}^N$, the corrections from the SU(3) breaking effects to their effective potentials become important.

We introduce the SU(3) breaking effects through two ways. Firstly, due to the differences between the constituent ($u$, $d$) quark masses and $s$ quark mass, the physical masses of the $\bar{D}^{(*)}_s$ and $\Omega_{cc}^{(*)}$ hadrons are heavier than that of the $\bar{D}^{(*)}$ and $\Xi_{cc}^{(*)}$ hadrons, respectively. We adopt their physical masses to partly introduce the SU(3) breaking effect.

Secondly, the effective potentials of the $\bar{T}_{cc}^f$/$P_{\psi c}^\Lambda$/$H_{\Omega_{ccc}c}^\Lambda$ states are induced from the exchanges of the isospin singlet and triplet light mesons, while the effective potentials of the $\bar{T}_{cc\bar{s}}^\theta/P_{\psi cs}^N$/$H_{\Omega_{ccc}cs}^N$ states are induced from the exchanges of the isospin singlet and doublet light mesons.
Thus, in the effective potentials of the $\bar{T}_{cc}^f$/$P_{\psi c}^\Lambda$/$H_{\Omega_{ccc}c}^\Lambda$ and $\bar{T}_{cc\bar{s}}^\theta/P_{\psi cs}^N$/$H_{\Omega_{ccc}cs}^N$ states, we introduce a SU(3) breaking factor $g_x$ to describe the different couplings of the contributions from the exchanges of isospin triplet and doublet light mesons, respectively.
\subsubsection{The existences of $\bar{T}_{cc\bar{s}}^\theta$ bound states}
The effective potentials of the $[\bar{D}_s\bar{D}^*]_1^{\frac{1}{2}}$ and $[\bar{D}^*\bar{D}^*_s]_1^{\frac{1}{2}}$ states have already been presented in Table \ref{matrixele}. As given in Table \ref{matrixele}, for the $[\bar{D}_s\bar{D}^*]_1^{\frac{1}{2}}$ state, we consider the mass difference between the $D_s^-\bar{D}^{*0}$ (3975.9 MeV) and $\bar{D}^0D_s^{*-}$ (3977.0 MeV) components and search for the bound state solution and quasi-bound state solution in the $E<E_{D_s^-\bar{D}^{*0}}$ and $E_{D_s^-\bar{D}^{*0}}<E<E_{\bar{D}^0D_s^{*-}}$ regions, respectively.

In the both scenarios, we run the $g_x$ value in the $[0,1]$ region and find that at small $g_x$ value, the attractive force is not enough to form the $[\bar{D}_s\bar{D}^*]_1^{\frac{1}{2}}$ bound state. The attractive force increases as the $g_x$ increases. When the $g_x$ is close to 1, the $[\bar{D}_s\bar{D}^*]_1^{\frac{1}{2}}$ state start to form bound state, its mass lies below the $D_s^-\bar{D}^{*0}$ threshold. As presented in Table \ref{matrixele}, when the $g_x$ is close to 1, the effective potential of the $[\bar{D}_s\bar{D}^*]_1^{\frac{1}{2}}$ state is almost identical to that of the $[D\bar{D}^*]_1^0$ state. Similar to the $[D\bar{D}^*]_1^0$ state, when the $g_x$ is close to 1, the binding energy of the $[\bar{D}_s\bar{D}^*]_1^{\frac{1}{2}}$ bound state is below and at the edge of the $D_s^-\bar{D}^{*0}$ threshold. However, since the $g_x$ value is further constrained by the mass of the $P_{\psi s}^\Lambda(4338)$, i.e., $g_x\leq 0.62$. 
To give a more serious conclusion on whether the $[\bar{D}_s\bar{D}^*]_1^{\frac{1}{2}}$ bound state can exist, we need to consider both the experimental uncertainties from the masses of the $P_{\psi}^N(4440)$ and $P_{\psi}^N(4457)$) states, and the $g_x$-dependence on the pole position of the $[\bar{D}_s\bar{D}^*]_1^{\frac{1}{2}}$ state.


\begin{figure}[htbp]
    \centering
    \includegraphics[width=1.0\linewidth]{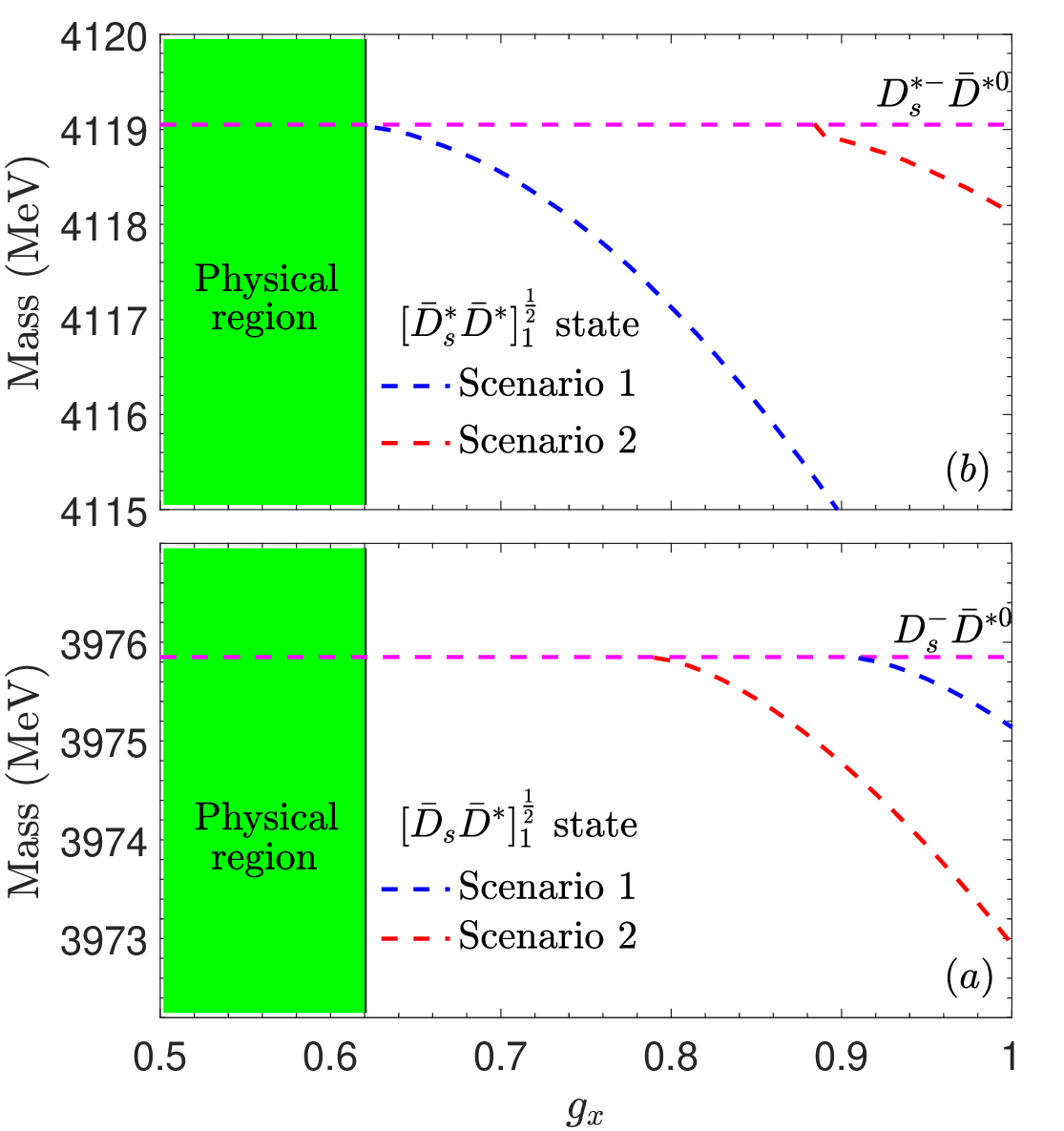}
    \caption{As the $g_x$ increases, we illustrate the \textit{lower limits} of the masses of the $[\bar{D}_s\bar{D}^*]_{1}^{\frac{1}{2}}$ and $[\bar{D}^*_s\bar{D}^*]_{1}^{\frac{1}{2}}$ states in (a) and (b), respectively. The results from the scenario 1 and scenario 2 are plotted with blue-dotted and red-dotted lines, respectively. The green area with $g_x\leq0.62$ is the parameter region that the $P_{\psi s}^\Lambda(4338)$ state can exist.}
    \label{Tccslimit}
\end{figure}

Now we proceed to include the experimental uncertainties from the masses of the $P_{\psi}^N(4440)$ and $P_{\psi}^N(4457)$ states. The scheme we give the lower and upper limits of our predictions has been discussed in detail in Sec. \ref{Parameters}, i.e., we combine the lower or upper limit of the mass of $P_{\psi}^N(4440)$ state with the lower or upper limit of the mass of $P_{\psi}^N(4457)$ state, we use these four sets masses as inputs to solve four sets of $\tilde{g}_s$ and $\tilde{g}_a$ coupling parameters correspondingly. Then we select the lower limit (with the maximum attractive force within the experimental error) and upper limit (with minimum attractive force within the experimental error) of the mass of the $[\bar{D}_s\bar{D}^*]_1^{\frac{1}{2}}$ state.

We illustrate the $g_x$-dependence of the lower limit for the bound state solution of the $[\bar{D}_s\bar{D}^*]_1^{\frac{1}{2}}$ state in Fig. \ref{Tccslimit} (a).  We use the blue-dotted and red-dotted lines to plot the results from  the scenario 1 and scenario 2, respectively. As can be seen from Fig. \ref{Tccslimit} (a), in the scenario 1 and scenario 2 cases, the lower limits of the masses of the $[\bar{D}_s\bar{D}^*]_1^{\frac{1}{2}}$ state are far away from the physical region of the SU(3) breaking factor $g_x$. The results from both scenarios show that the strangeness partner of the $T_{cc}^f(3875)$ can not exist after including the SU(3) breaking effect.


Similarly, in both scenarios, we plot the $g_x$-dependences of the lower limits for the bound state solutions of the $[\bar{D}^*\bar{D}^*_s]_1^{\frac{1}{2}}$ state in Fig. \ref{Tccslimit} (b). As illustrated in Fig. \ref{Tccslimit} (b), in the scenario 1, the lower limit of the mass of the $[\bar{D}^*\bar{D}^*_s]_1^{\frac{1}{2}}$ bound state labeled with blue-dotted line has a very tiny overlap with the physical $g_x$ region. We can not entirely exclude the existence of the $[\bar{D}^*\bar{D}^*_s]_1^{\frac{1}{2}}$ bound state, but our result prefers the conclusion that the $[\bar{D}^*\bar{D}^*_s]_1^{\frac{1}{2}}$ does not exist.

In the scenario 2, the lower limit of the pole position for the $[\bar{D}^*\bar{D}^*_s]_1^{\frac{1}{2}}$ bound state is labeled with red-dotted line, as presented Fig. \ref{Tccslimit} (b), our result shows that the lower limit of the pole position for the $[\bar{D}^*\bar{D}^*_s]_1^{\frac{1}{2}}$ bound state is far away from the physical region of the SU(3) breaking factor $g_x$, in this scenario, the existence of the $[\bar{D}^*\bar{D}^*_s]_1^{\frac{1}{2}}$ bound state can be excluded. Or vise versa, the result of the $[\bar{D}^*\bar{D}^*_s]_1^{\frac{1}{2}}$ state in the scenario 2 suggests that once the $J^P$ numbers of the $P_{\psi}^N(4440)$ and $P_{\psi}^N(4457)$ were measured to be $\frac{3}{2}^-$ and $\frac{1}{2}^-$, respectively, then we can conclude that the $[\bar{D}^*\bar{D}^*_s]_1^{\frac{1}{2}}$ bound state do not exist.
\subsubsection{The existences of $P_{\psi cs}^N$ bound states}
Now we proceed to discuss the existences of the $P_{\psi cs}^N$ bound states. As presented in Table \ref{thresholds}, due to the large constituent strange quark mass, for a triple-charm baryon-meson (B-M) system, the strange quark belongs to the double-charm baryon or charmed meson will lead to different B-M systems. Thus, we investigate the possible $P_{\psi cs}^N$ bound states via a mixing scheme. We mix the $(ccn)-(\bar{c}s)$ and $(ccs)-(\bar{c}n)$ components to derive the effective potentials of the considered $P_{\psi cs}^N$ states.

As listed in Table \ref{matrixele}, we need to consider the 7 pairs of B-M systems, i.e, the ($[\Xi_{cc}\bar{D}_s]_{\frac{1}{2}}$, $[\Omega_{cc}\bar{D}]_{\frac{1}{2}}$), ($[\Xi_{cc}\bar{D}^*_s]_{\frac{1}{2}}$, $[\Omega_{cc}\bar{D}^*]_{\frac{1}{2}}$), ($[\Xi^*_{cc}\bar{D}^*_s]_{\frac{1}{2}}$, $[\Omega^*_{cc}\bar{D}^*]_{\frac{1}{2}}$), ($[\Xi^*_{cc}\bar{D}_s]_{\frac{3}{2}}$, $[\Omega^*_{cc}\bar{D}]_{\frac{3}{2}}$), ($[\Xi_{cc}\bar{D}^*_s]_{\frac{3}{2}}$, $[\Omega_{cc}\bar{D}^*]_{\frac{3}{2}}$), ($[\Xi^*_{cc}\bar{D}^*_s]_{\frac{3}{2}}$, $[\Omega^*_{cc}\bar{D}^*]_{\frac{3}{2}}$), and ($[\Xi^*_{cc}\bar{D}^*_s]_{\frac{5}{2}}$, $[\Omega^*_{cc}\bar{D}^*]_{\frac{5}{2}}$) systems. Correspondingly, their SU(3) flavor partners in the isospin symmetry limit are the $[\Xi_{cc}^*\bar{D}]_{\frac{1}{2}}^0$, $[\Xi_{cc}\bar{D}^*]_{\frac{1}{2}}^0$, $[\Xi_{cc}^*\bar{D}^*]_{\frac{1}{2}}^0$, $[\Xi_{cc}^*\bar{D}]_{\frac{3}{2}}^0$, $[\Xi_{cc}\bar{D}^*]_{\frac{3}{2}}^0$, $[\Xi_{cc}^*\bar{D}]_{\frac{3}{2}}^0$, and $[\Xi_{cc}^*\bar{D}^*]_{\frac{5}{2}}^0$, respectively. Here, we need to emphasis that these non-strange $P_{\psi c}^\Lambda$ states all have bound state solutions as discussed in Sec. \ref{nonstrange}.

From Table \ref{thresholds}, the threshold of the $\Xi^{(*)}_{cc}\bar{D}_s^{(*)}$ system is lower than that of the $\Omega_{cc}^{(*)}\bar{D}^{(*)}$ threshold, thus, in each pair of B-M system, we search for the possible bound state solution and quasi-bound state solution in the energy region $E<E_{\Xi^{(*)}_{cc}\bar{D}_s^{(*)}}$ and $E_{\Xi^{(*)}_{cc}\bar{D}_s^{(*)}}<E<E_{\Omega_{cc}^{(*)}\bar{D}^{(*)}}$, respectively. We find that the bound state below the $\Xi^{(*)}_{cc}\bar{D}_s^{(*)}$ threshold may exist in the SU(3) limit with $g_x=1$, but there is no quasi-bound state in the energy region between the $\Xi^{(*)}_{cc}\bar{D}_s^{(*)}$ and $\Omega_{cc}^{(*)}\bar{D}^{(*)}$ thresholds.

We test the $g_x$-dependences of the bound state solutions that are below the corresponding $\Xi_{cc}^{(*)}\bar{D}_s^{(*)}$ thresholds for the considered 7 pairs of B-M systems. However, unlike the $P_{\psi c}^\Lambda$ states, we find that all these 7 pairs of B-M systems can hardly form bound states when the $g_x\leq 0.62$, i.e., in the physical $g_x$ region that the $P_{\psi s}^{\Lambda}(4338)$ can exist. Especially, according to our calculation, among these 7 pairs of B-M systems, the ($[\Xi^*_{cc}\bar{D}^*_s]_{\frac{1}{2}}$, $[\Omega^*_{cc}\bar{D}^*]_{\frac{1}{2}}$) system in the scenario 1 has the most attractive force. However, even for this extreme case, our result still prefers the non-existence of the $[\Xi^*_{cc}\bar{D}^*_s]^{\frac{1}{2}}_{\frac{1}{2}}$ bound state.

\begin{figure}[htbp]
    \centering
    \includegraphics[width=1.0\linewidth]{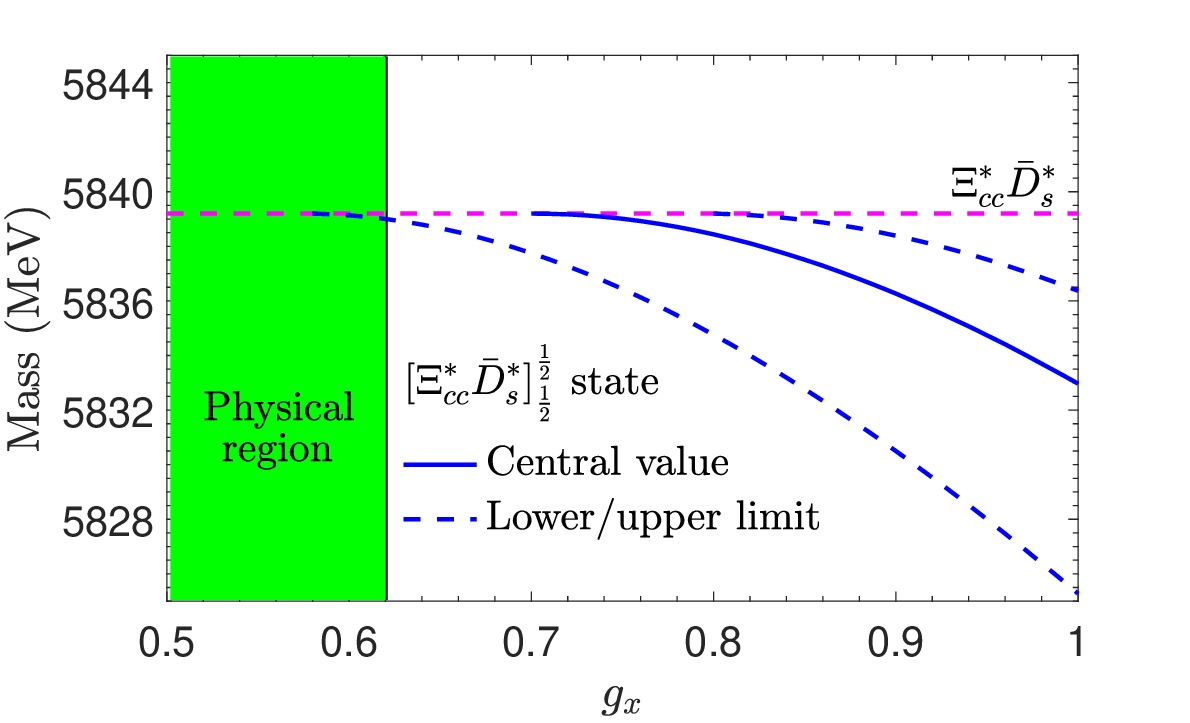}
    \caption{From the left to right are the $g_x$-dependences of \textit{lower limit} (blue-dotted line), \textit{central value} (blue solid line), and \textit{upper limit} (blue-dotted line) of the bound state solutions for the $[\Xi_{cc}^{*}\bar{D}_s^*]_{\frac{1}{2}}^{\frac{1}{2}}$ state with the inputs from the scenario 1. The green area with $g_x\leq0.62$ is the parameter region that the $P_{\psi s}^\Lambda(4338)$ can exist.}
    \label{Pcccslimit}
\end{figure}

In Fig. \ref{Pcccslimit}, we plot the $g_x$-dependence of the $[\Xi_{cc}^{*}\bar{D}_s^*]_{\frac{1}{2}}^{\frac{1}{2}}$ bound state solution calculated from the effective potential of ($[\Xi^*_{cc}\bar{D}^*_s]_{\frac{1}{2}}$, $[\Omega^*_{cc}\bar{D}^*]_{\frac{1}{2}}$) system. In Fig. \ref{Pcccslimit}, the blue-dotted lines on the left and right are the lower and upper limits of the bound state solutions for the $[\Xi_{cc}^{*}\bar{D}_s^*]_{\frac{1}{2}}^{\frac{1}{2}}$ state. The blue solid line is obtained with the inputs from the central values of the masses of $P_{\psi}^N(4440)$ and $P_{\psi}^N(4457)$ states.

As can be seen from Fig. \ref{Pcccslimit}, the lower limit of the mass of the $[\Xi_{cc}^{*}\bar{D}_s^*]_{\frac{1}{2}}^{\frac{1}{2}}$ bound state has a tiny overlap with the physical $g_x$ region. From this result, we can conclude that the $[\Xi_{cc}^{*}\bar{D}_s^*]_{\frac{1}{2}}^{\frac{1}{2}}$ bound state can hardly exist. We do not further plot the $g_x$ dependences for the rest of the 6 pairs of the ($[\Xi^{(*)}_{cc}\bar{D}^{(*)}_s]$, $[\Omega^{(*)}_{cc}\bar{D}^{(*)}]$) systems, since the attractive forces in these 6 systems in both scenarios are all weaker than that of the discussed
($[\Xi^*_{cc}\bar{D}^*_s]_{\frac{1}{2}}$, $[\Omega^*_{cc}\bar{D}^*]_{\frac{1}{2}}$) system with the parameters from scenario 1 and thus can not form bound states. Collectively, there exist seven bound states in the $P_{\psi c}^\Lambda$ system, while all their SU(3) strange $P_{\psi cs}^N$ partners can hardly exist. This conclusion is very similar to the $\bar{T}_{cc}^f$ and $\bar{T}_{cc\bar{s}}^\theta$ case.

The significant differences between the $P_{\psi c}^\Lambda$ and $P_{\psi cs}^N$ systems are mainly attributed to the SU(3) breaking effects. Note that in our framework, the SU(3) breaking effects are mainly introduced from two sources. Firstly, the threshold of the $\Xi^{(*)}_{cc}\bar{D}^{(*)}$ is different from that of the $\Xi_{cc}^{(*)}\bar{D}^{(*)}_s$ or $\Omega_{cc}^{(*)}\bar{D}^{(*)}$ threshold due to the large constituent strange quark mass. Secondly, the effective potentials of the $P_{\psi c}^\Lambda$ states consist of the interactions introduced from the exchanges of the isospin singlet and triplet light mesons, while the effective potentials of the $P_{\psi cs}^N$ states consist of the interactions introduced from the exchanges of the isospin singlet and doublet light mesons, thus, comparing to the contributions from the exchanges of the isospin triplet light mesons, the contributions from the exchanges of the isospin doublet light mesons are expected to be suppressed by the factor $1/m_{\text{ex}}^2$.

Here, we present more discussions to clarify how these two SU(3) breaking sources influence the formation of the $\Xi^{(*)}_{cc}\bar{D}^{(*)}_s$ bound state. The possible $\Xi^{(*)}_{cc}\bar{D}^{(*)}_s$ bound state is investigated from the effective potential of the ($\Xi_{cc}^{(*)}\bar{D}_s^{(*)}$, $\Omega_{cc}^{(*)}\bar{D}^{(*)}$) system listed in Table \ref{matrixele}, to construct the direct relation between the $\Xi^{(*)}_{cc}\bar{D}^{(*)}$ and $\Xi_{cc}^{(*)}\bar{D}^{(*)}_s$ states, we demonstrate the following analogy
\begin{eqnarray}
&&V_{\Xi^{(*)}_{cc}\bar{D}^{(*)}}\nonumber\\&=&\left(
                                      \begin{array}{cc}
                                        v^{\Xi_{cc}^{(*)++}D^{(*)-}\rightarrow \Xi_{cc}^{(*)++}D^{(*)-}} & v^{\Xi_{cc}^{(*)++}D^{(*)-}\rightarrow \Xi_{cc}^{(*)+}\bar{D}^{(*)0}} \\
                                        v^{\Xi_{cc}^{(*)+}\bar{D}^{(*)0}\rightarrow \Xi_{cc}^{(*)++}D^{(*)-}} & v^{\Xi_{cc}^{(*)+}\bar{D}^{(*)0}\rightarrow \Xi_{cc}^{(*)+}\bar{D}^{(*)0}} \\
                                      \end{array}
                                    \right),\nonumber\label{SU2BM}\\
\end{eqnarray}
\begin{eqnarray}
&&V_{(\Xi^{(*)}_{cc}\bar{D}_s^{(*)},\Omega_{cc}^{(*)}\bar{D}^{(*)})}\nonumber\\&=&\left(
                                      \begin{array}{cc}
                                        v^{\Xi_{cc}^{(*)++}D_s^{(*)-}\rightarrow \Xi_{cc}^{(*)++}D_s^{(*)-}} & v^{\Xi_{cc}^{(*)++}D_s^{(*)-}\rightarrow \Omega_{cc}^{(*)+}\bar{D}^{(*)0}} \\
                                        v^{\Omega_{cc}^{(*)+}\bar{D}^{(*)0}\rightarrow \Xi_{cc}^{(*)++}D_s^{(*)-}} & v^{\Omega_{cc}^{(*)+}\bar{D}^{(*)0}\rightarrow \Omega_{cc}^{(*)+}\bar{D}^{(*)0}} \\
                                      \end{array}
                                    \right).\nonumber\label{SU3BM}\\
\end{eqnarray}
Here, the $\Xi_{cc}^{(*)++}D^{(*)-}$ and $\Xi_{cc}^{(*)+}\bar{D}^{(*)0}$ components in Eq. (\ref{SU2BM}) correspond to the $\Xi_{cc}^{(*)++}D_s^{(*)-}$ and $\Omega_{cc}^{(*)+}\bar{D}^{(*)0}$ components in Eq. (\ref{SU3BM}), respectively. Note that the magnitude of the mass correction from the SU(3) breaking is much larger than that of the isospin breaking, thus, the mass gap between the $\Xi_{cc}^{*++}D^{*-}$ and $\Xi_{cc}^{*+}\bar{D}^{*0}$ components are much smaller than that of the $\Xi_{cc}^{*++}D^{*-}_s$ and $\Omega_{cc}^{*+}\bar{D}^{*0}$ components. Thus, comparing to the off-diagonal $\Xi_{cc}^{(*)++}D^{(*)-}-\Xi_{cc}^{(*)+}\bar{D}^{(*)0}$ coupling, the off-diagonal $\Xi_{cc}^{(*)++}D_s^{(*)-}-\Omega_{cc}^{(*)+}\bar{D}^{(*)0}$ coupling is suppressed by the large mass gap between the $\Xi^{(*)++}_{cc}\bar{D}_s^{(*)-}$ and $\Omega_{cc}^{(*)+}\bar{D}^{(*)0}$ components. Thus, the SU(3) breaking effect induced from the mass gap of the $\Xi_{cc}^{(*)++}D_s^{(*)-}-\Omega_{cc}^{(*)+}\bar{D}^{(*)0}$ system will suppress the attractive force in the attractive $\Xi^{(*)}_{cc}\bar{D}_s^{(*)}$ channel.

On the other hand, since the off-diagonal $\Xi_{cc}^{*++}D_s^{*-}-\Omega_{cc}^{*+}\bar{D}^{*0}$ channel can exchange strange isospin doublet light mesons, comparing to the off-diagonal $\Xi_{cc}^{*++}D^{*-}-\Xi_{cc}^{*+}\bar{D}^{*0}$ channel, the $\Xi_{cc}^{*++}D_s^{*-}-\Omega_{cc}^{*+}\bar{D}^{*0}$ coupling should be further suppressed by the mass of the exchanged strange light mesons. The double-suppression from these two SU(3) breaking sources are the main reasons that the $\Xi_{cc}^{(*)}\bar{D}_s^{(*)}$ bound state can hardly exist.


\subsubsection{The existences of $H_{\Omega_{ccc}cs}^N$ bound states}
Now we discuss the possible $H_{\Omega_{ccc}cs}^N$ bound states. As presented in Table \ref{matrixele}, we search for the possible bound states in the $\Xi_{cc}\Omega_{cc}$ and $\Xi_{cc}^*\Omega_{cc}^*$ systems via solving the single-channel LS equation. But for the ($\Xi_{cc}\Omega^*_{cc}$, $\Xi_{cc}^*\Omega_{cc}$) system, we perform a two-channel calculation and search for the bound state solution and quasi-bound state solution in the energy regions $E<E_{\Xi_{cc}\Omega^*_{cc}}$ and $E_{\Xi_{cc}\Omega_{cc}^*}<E<E_{\Xi^*_{cc}\Omega_{cc}}$, respectively. We find that there are no quasi-bound states in the $E_{\Xi_{cc}\Omega_{cc}^*}<E<E_{\Xi^*_{cc}\Omega_{cc}}$ region, but in the $E<E_{\Xi_{cc}\Omega^*_{cc}}$ region, the $\Xi_{cc}\Omega_{cc}^*$ bound states may exist.

\begin{figure}[htbp]
    \centering
    \includegraphics[width=1.0\linewidth]{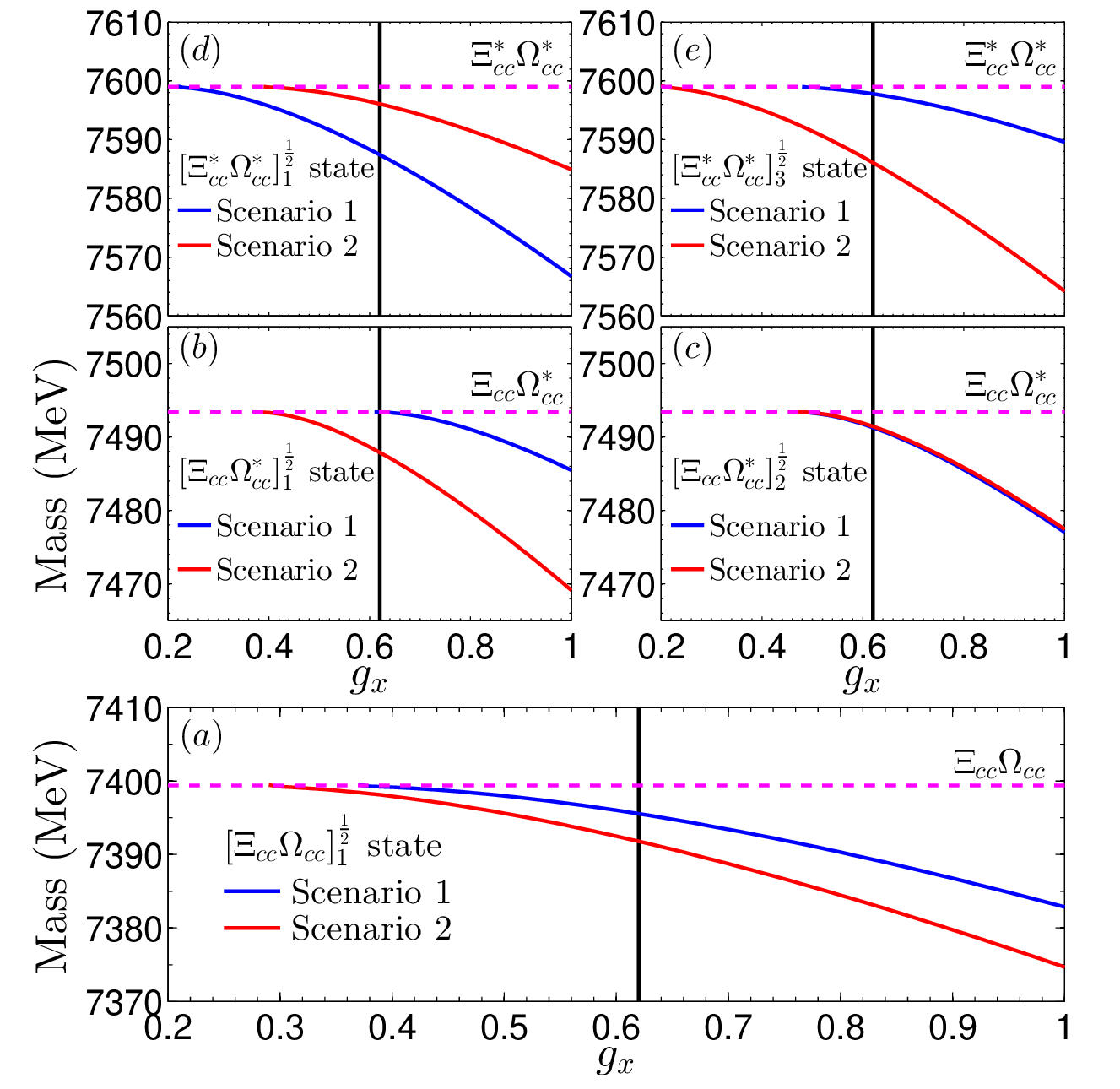}
    \caption{The $g_x$-dependences of the bound state solutions for the (a)-(e): $[\Xi_{cc}\Omega_{cc}]_1^{\frac{1}{2}}$, $[\Xi_{cc}\Omega_{cc}^*]_1^{\frac{1}{2}}$, $[\Xi_{cc}\Omega_{cc}^*]_2^{\frac{1}{2}}$, $[\Xi_{cc}^*\Omega_{cc}^*]_1^{\frac{1}{2}}$, and $[\Xi_{cc}^*\Omega_{cc}^*]_3^{\frac{1}{2}}$ states. The bound state solutions are calculated with the parameters solved from the experimental \textit{central values} of the masses of the $P_{\psi}^N(4440)$ and $P_{\psi}^N(4457)$ states. The blue solid and red solid lines are used to label the results calculated from the scenario 1 and scenario 2.}
    \label{HOmegaccccs}
\end{figure}

We investigate the existences of the $H_{\Omega_{ccc}cs}^N$ bound states by including the SU(3) breaking effects. In Fig. \ref{HOmegaccccs}, we plot the $g_x$-dependences of the bound state solutions for the considered $H_{\Omega_{ccc}cs}^N$ states.  We use the parameters solved from the experimental central values of the masses of the $P_{\psi}^N(4440)$ and $P_{\psi}^N(4457)$ states to calculate the possible bound states in the $H_{\Omega_{ccc}cs}^N$ system. Then we use the blue and red solid lines to label the results calculated with the inputs from the scenario 1 and scenario 2, respectively. As illustrated in Fig. \ref{HOmegaccccs} (a)-(e), we find that at $g_x=0.62$, the $[\Xi_{cc}\Omega_{cc}]_1^{\frac{1}{2}}$, $[\Xi_{cc}\Omega_{cc}^*]_1^{\frac{1}{2}}$, $[\Xi_{cc}\Omega_{cc}^*]_2^{\frac{1}{2}}$, $[\Xi_{cc}^*\Omega_{cc}^*]_1^{\frac{1}{2}}$, and $[\Xi_{cc}^*\Omega_{cc}^*]_3^{\frac{1}{2}}$ all have bound state solutions in the both scenarios.

Note that comparing to the states in the $\bar{T}_{cc\bar{s}}^N$ and $P_{\psi cs}^{N}$ systems, the states in the $H_{\Omega_{ccc}cs}^N$ system have larger reduced masses. Thus, our results for the $H_{\Omega_{ccc}cs}^N$ bound states show the important role of the reduced masses in stabilizing heavy flavor molecular states.

Besides, we can also obtain the different mass arrangements of the $H_{\Omega_{ccc}cs}^N$ bound states in the scenario 1 and scenario 2. Explicitly, in the scenario 1, as plotted in Fig. \ref{HOmegaccccs} (b) and (c), the mass of the bound state that is mainly composed of the $\Xi_{cc}\Omega_{cc}^*$ decreases as the total angular momentum increases, while as plotted in Fig. \ref{HOmegaccccs} (d) and (e), the mass of the bound state that is mainly composed of the $\Xi_{cc}^*\Omega_{cc}^*$ increases as the total angular momentum increases. On the contrary, the above tendencies are opposite in the scenario 2 case.

\section{Summary}\label{summary}

The special formation mechanism of the molecular states composed of two heavy-light hadrons motivates us to introduce a symmetric framework to relate the interactions of different di-hadron systems that are composed of two heavy-light hadrons.

We suggest that the SU(3) flavor symmetry together with the SU(2) spin symmetry can be used to relate the di-hadron systems with different light quark components, i.e.,
\begin{eqnarray}
&P_{\psi}^N(cnn)(\bar{c}n)&\leftrightarrow \bar{T}_{cc}^f (\bar{c}n)(\bar{c}n) ,\\
&P_{\psi s}^\Lambda\begin{cases}
    (cns)(\bar{c}n)\\
    (cnn)(\bar{c}s)
    \end{cases}&
    \leftrightarrow \bar{T}_{cc\bar{s}}^\theta (\bar{c}n)(\bar{c}s),
\end{eqnarray}
while the HDAS together with the SU(2) spin symmetry can be used to relate the di-hadron systems with different heavy quark components, i.e.,
\begin{eqnarray}
\bar{T}_{cc}^f\,(\bar{c}n)(\bar{c}n)\leftrightarrow &P_{\psi c}^\Lambda\,(ccn)(\bar{c}n)&\leftrightarrow H_{\Omega_{ccc}c}^\Lambda\,(ccn)(ccn),\nonumber\\
\bar{T}_{cc\bar{s}}^\theta\,(\bar{c}n)(\bar{c}s)\leftrightarrow &P_{\psi cs}^N\,\begin{cases}
    (ccs)(\bar{c}n)\\
    (ccn)(\bar{c}s)
    \end{cases}&\leftrightarrow H_{\Omega_{ccc}cs}^N\,(ccn)(ccs) . \nonumber
\end{eqnarray}

This work is devoted to give a possible unified description to the mass spectra of $\bar{T}^f_{cc}$/$\bar{T}_{ccs}^\theta$, $P_{\psi c}^\Lambda$/$P_{\psi cs}^N$, and $H_{\Omega_{ccc}c}^\Lambda$/$H_{\Omega_{ccc}cs}^N$ systems. We introduce a contact lagrangian possessing the SU(3) flavor and SU(2) spin symmetries to describe the interactions of the considered di-hadron systems. Then we include the SU(3) breaking effects from two sources. Firstly, the large violation of SU(3) symmetry is reflected by the physical masses of strange double-charm baryons and single-charm mesons, we adopt their physical masses to perform our calculations. Secondly, the $\bar{T}_{cc\bar{s}}^\theta$/$P_{\psi cs}^N$/$H_{\Omega_{ccc}cs}^N$ systems can exchange the isospin doublet light mesons, which should be suppressed by the large masses of exchanged strange light mesons, we describe this suppression by introducing an SU(3) breaking factor $g_x$.

We introduce four parameters in our model, the $\Lambda$, $\tilde{g}_s$, $\tilde{g}_a$, and $g_x$. Here, the cutoff $\Lambda$ in our dipole form factor is fixed at 1.0 GeV throughout this work. The $J^P$ quantum numbers of the $P_{\psi}^N(4440)$ and $P_{\psi}^N(4457)$ are assumed in the two scenarios, correspondingly, the parameters $\tilde{g}_s$ and $\tilde{g}_a$ are solved in the both scenarios. Then we further constrain the SU(3) breaking factor $g_x$ by considering the mass of the observed $P_{\psi s}^\Lambda(4338)$ state.

In both scenarios, we calculate the mass spectra of the $\bar{T}_{cc}^f$/$P_{\psi c}^\Lambda$/$H_{\Omega_{ccc}c}^\Lambda$ bound states. We demonstrate that according to the flavor-spin symmetry, the measurements of the $J^P$ quantum numbers for the $P_{\psi}^N(4440)$ and $P_{\psi}^N(4457)$ states would provide important information to the binding energies of the bound states in the $\bar{T}_{cc}^f$ system. We also present that these two scenarios will lead to different mass arrangements in the $P_{\psi c}^\Lambda$ and $H_{\Omega_{ccc}c}^\Lambda$ mass spectra. All these discussed signatures can be used to test the flavor-spin symmetry among the interactions of the heavy flavor molecule community.

We also investigate the existences of the $\bar{T}_{cc\bar{s}}^\theta$/$P_{\psi cs}^N$/$H_{\Omega_{ccc}cs}^N$ bound states. By considering the SU(3) breaking effects from the two aforementioned sources, we demonstrate that for the $\bar{T}_{cc\bar{s}}^\theta$/$P_{\psi cs}^N$/$H_{\Omega_{ccc}cs}^N$ states that have attractive effective potentials, these two sources of SU(3) breaking effects both suppress the attractive forces in the $\bar{T}_{cc\bar{s}}^\theta$/$P_{\psi cs}^N$/$H_{\Omega_{ccc}cs}^N$ states. Thus, we find that unlike the $\bar{T}_{cc}^f$/$P_{\psi c}^\Lambda$ systems, the states in the $\bar{T}_{cc\bar{s}}^\theta$/$P_{\psi cs}^N$ systems can hardly form bound states. However, due to the large reduced masses, the states in the $H_{\Omega_{ccc}cs}^N$ systems can still form bound states although the double-suppression from the SU(3) breaking effects will significantly decrease the absolute values of their binding energies.

Finally, we need to emphasis that the $\bar{T}^f_{cc}$/$\bar{T}_{cc\bar{s}}^\theta$, $P_{\psi c}^\Lambda$/$P_{\psi cs}^N$, $H_{\Omega_{ccc}c}^\Lambda$/$H_{\Omega_{ccc}cs}^N$ systems discussed in this work and the $P_{\psi}^N$/$P_{\psi s}^\Lambda$, $H_{\Omega_{ccc}}^N$/$H_{\Omega_{ccc}s}^\Lambda$ systems discussed in our previous work \cite{Chen:2024tuu} are calculated within the same framework and identical parameters, this allow us to build direct relations from one di-hadron system to another di-hadron system through the SU(3) flavor symmetry or HDAS symmetry together with the SU(2) spin symmetry. Thus, the experimental measurements from one of the above systems may give valuable informations to the rest of discussed systems. We hope that some of our predictions could be confirmed by the lattice QCD simulations or the BESIII, LHCb, and BELLE II collaborations in the future.
\section*{Acknowledgments}
Kan Chen want to thank Zi-Yang Lin for helpful discussion. This work is supported by the National Natural Science
Foundation of China under Grants No. 12305090, 12105072.
Bo Wang is also supported by the Start-up Funds for Young Talents of Hebei University (No. 521100221021).

\end{document}